\documentclass{article}

\usepackage{arxiv}

\usepackage[utf8]{inputenc} 
\usepackage[T1]{fontenc}    
\usepackage{hyperref}       
\usepackage{url}            
\usepackage{booktabs}       
\usepackage{amsfonts}       
\usepackage{nicefrac}       
\usepackage{microtype}      
\usepackage{lipsum}		
\usepackage{graphicx}
\usepackage{natbib}
\usepackage{doi}

\usepackage{color}
\usepackage{amsmath}
\usepackage{amsthm}
\usepackage{amsfonts}
\usepackage{natbib}
\usepackage[noend]{algpseudocode}
\usepackage{tikz}
\usepackage{xcolor}
\usepackage{color, colortbl}
\usepackage{amssymb}
\usepackage[nodisplayskipstretch]{setspace}
\usepackage{subfig}
\usepackage{xcolor}
\usepackage[linesnumbered,ruled,vlined]{algorithm2e}

\SetCommentSty{mycommfont}
\usepackage[noend]{algpseudocode}
\SetKwInput{KwInput}{Input}                
\SetKwInput{KwOutput}{Output}              
\SetKwInput{KwParameter}{Parameter} 

\usetikzlibrary{shadows}
\tikzset{
diagonal fill/.style 2 args={fill=#2, path picture={
\fill[#1, sharp corners] (path picture bounding box.south west) -|
                         (path picture bounding box.north east) -- cycle;}},
reversed diagonal fill/.style 2 args={fill=#2, path picture={
\fill[#1, sharp corners] (path picture bounding box.north west) |- 
                         (path picture bounding box.south east) -- cycle;}}
}

\usetikzlibrary{bayesnet}
\usetikzlibrary{arrows}
\usetikzlibrary{calc}
\usetikzlibrary{shapes,decorations,arrows,calc,arrows.meta,fit,positioning, automata}
\tikzset{
    -Latex,auto,node distance =1 cm and 1 cm,semithick,
    state/.style ={ellipse, draw, minimum width = 0.7 cm},
    point/.style = {circle, draw, inner sep=0.04cm,fill,node contents={}},
    bidirected/.style={Latex-Latex,dashed},
    el/.style = {inner sep=2pt, align=left, sloped}
}
\algnewcommand{\LineComment}[1]{ \(\blacktriangleright\) \emph{\color{brown} #1}}

\def\eqref#1{Eq.~(\ref{#1})}
\def\figref#1{Fig.~\ref{#1}}
\def\secref#1{Sec.~\ref{#1}}
\def\algref#1{Algo~\ref{#1}}

\def\lineref#1{line~\ref{#1}}
\def\tableref#1{Table~\ref{#1}}

\title{Experimental design for causal query estimation in partially observed biomolecular networks}

\author{ \href{https://orcid.org/0000-0002-6554-9083}{\includegraphics[scale=0.06]{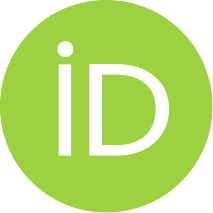}\hspace{1mm}Sara ~Mohammad-Taheri}\thanks{Corresponding author.} \\
	Khoury College of Computer Sciences\\
	Northeastern University\\
	Boston, MA 02115 \\
	\texttt{mohammadtaheri.s@northeastern.edu} \\
	\And
	\href{https://orcid.org/0000-0003-0002-4259}{\includegraphics[scale=0.06]{orcid.pdf}\hspace{1mm}Vartika ~Tewari} \\
	Khoury College of Computer Sciences\\
	Northeastern University\\
	Boston, MA 02115 \\
	\texttt{tewari.v@northeastern.edu} \\
	\And
	\href{https://orcid.org/0000-0000-0000-0000}{\includegraphics[scale=0.06]{orcid.pdf}\hspace{1mm}Rohan ~Kapre} \\
	Khoury College of Computer Sciences\\
	Northeastern University\\
	Boston, MA 02115 \\
	\texttt{kapre.r@northeastern.edu} \\
	\And
	\href{https://orcid.org/0000-0000-0000-0000}{\includegraphics[scale=0.06]{orcid.pdf}\hspace{1mm}Karen ~Sachs} \\
	Next Generation Analytics\\
	Palo Alto, CA\\
	\texttt{sachskaren@gmail.com} \\
	\And
	\href{https://orcid.org/0000-0000-0000-0000}{\includegraphics[scale=0.06]{orcid.pdf}\hspace{1mm}Charles Tapley ~Hoyt} \\
	Laboratory of Systems Pharmacology\\
	Harvard Medical School\\
	Boston, MA\\
	\texttt{cthoyt@gmail.com} \\
	\And
	\href{https://orcid.org/0000-0002-7276-9009}{\includegraphics[scale=0.06]{orcid.pdf}\hspace{1mm}Jeremy ~Zucker} \\
	Pacific Northwest National Laboratory\\
	Richland, WA 99354\\
	\texttt{jeremy.zucker@pnnl.gov} \\
	\And
	\href{https://orcid.org/0000-0003-1728-1104}{\includegraphics[scale=0.06]{orcid.pdf}\hspace{1mm}Olga ~Vitek} \\
	\thanks{Corresponding author.} \\
	Khoury College of Computer Sciences\\
	Northeastern University\\
	Boston, MA 02115 \\
	\texttt{o.vitek@northeastern.edu} \\
}

\renewcommand{\shorttitle}{\textit{arXiv} Template}

\hypersetup{
pdftitle={A template for the arxiv style},
pdfsubject={q-bio.NC, q-bio.QM},
pdfauthor={David S.~Hippocampus, Elias D.~Striatum},
pdfkeywords={First keyword, Second keyword, More},
}

\begin{document}
\maketitle

\begin{abstract}
Estimating a causal query from observational data is an essential task in the analysis of biomolecular networks.
Estimation takes as input a network topology, a query estimation method, and observational measurements on the network variables.
However, estimations involving many variables can be experimentally expensive, and computationally intractable.  
Moreover, using the full set of variables can be detrimental, leading to bias, or increasing the variance in the estimation.
Therefore, designing an experiment based on a well-chosen subset of network components can increase estimation accuracy, and reduce experimental and computational costs.
We propose a simulation-based algorithm for selecting sub-networks that support unbiased estimators of the causal query under a constraint of cost, ranked with respect to the variance of the estimators.
The simulations are constructed based on historical experimental data, or based on known properties of the biological system. 
Three case studies demonstrated the effectiveness of well-chosen network subsets for estimating causal queries from observational data.
All the case studies are reproducible and available at \url{https://github.com/srtaheri/Simplified_LVM}.
\end{abstract}


\section{Introduction}

Causal query estimation \cite{pearl2009causality} is an essential task in analysis of biomolecular networks.
Queries of the form \emph{``When we intervene on $X$, what is the effect on its descendent $Y$?"} provide insights into the function of biomolecular systems, enable medical decision making, and help develop drugs. Frequently, applying an intervention and collecting interventional data is technologically challenging or expensive. Therefore, methods have been proposed for estimating causal queries from observational data without interventions \cite{jung2020learning, jung2020estimating, bhattacharya2020semiparametric}.

Causal query estimation from observational data relies on a known biomolecular network, i.e. a graph where variables (nodes) are signaling proteins, genes, transcripts or metabolites, and directed edges are previously established causal regulatory relationships. 
Such networks are available in a variety of knowledge bases such as INDRA \cite{bachman2022automated}, Reactome \cite{gillespie2022reactome}, and Omnipath \cite{turei2016omnipath}.
Knowledge-based biomolecular networks are often large, and estimation of causal queries with the entire network can be computationally intractable. 
Perhaps counterintuitively, using the full set of variables can in fact be harmful, and lead to bias, or increase variance \cite{cinelli2020crash}.
Therefore, the experimental design should measure a well-chosen subset of variables that minimize the bias and the variance of a causal query estimator. 

Additionally, biomolecular measurements incur highly varying costs. 
For example, while developing a new antibody for flow cytometry experiments is expensive, quantifying a protein with an existing antibody is far cheaper. In mass spectrometry, measurements of low-abundance or modified proteins may require additional steps and incur further costs.
Subsets of the transcriptome may be quantified at lower cost than RNA sequencing, using other technologies such as qPCR. 
Thus, the constraints of measurement costs are an important consideration when designing an experiment.

Existing approaches for selecting a subset of variables for causal query estimation rely on graphical criteria to determine optimal adjustment sets.
For example, the adjustment sets include confounders, which, if left unaccounted for, induce spurious correlations.
While such criteria are clear for models with no latent variables~\cite{rotnitzky2020efficient, henckel2019graphical}, they have only been established in very restrictive situations for models where latent variables exist~\cite{runge2021necessary}.
Moreover, such adjustment sets typically overlook useful variables such as mediators (connecting the cause to the effect via a directed path), and do not consider experimental cost.


In this manuscript, we propose a more comprehensive approach for finding optimal sub-networks, to be quantified in a subsequent observational study, with the objective of minimizing the bias and variance of a causal query estimator under a constraint of measurement cost.

Unlike existing methods that determine optimal adjustment sets based on graphical criteria, we use simulations. 
Simulations have a long and successful history in computational biology, e.g. single-cell RNA-seq data simulation \cite{cao2021benchmark, marouf2020realistic,sun2021scdesign2}, stochastic simulation of biochemical reaction systems \cite{marchetti2017stochastic}, high-throughput sequencing data simulations (ReSeq)~\cite{schmeing2021reseq}, and genetic data simulators \cite{peng2015genetic}.
Simulations have been used to model true biological systems accurately, robustly and reproducibly~\cite{marouf2020realistic}, in particular in situations with unattainable ground truth~\cite{sun2021scdesign2,cao2021benchmark}. They also help in multi-criteria decision making by examining trade-offs~\cite{dunke2021simulation}.    

The proposed approach effectively explores subsets of the variables in the biomolecular network. 
It uses theoretical considerations, simulation models and sensitivity analysis to characterise the bias and variance of causal query estimators in the presence of latent variables. 
The simulation-based representation of system variability, and the exploration of broader subsets of variables, allow us to overcome the limitations of techniques that only rely on the adjustment sets. 
We demonstrate the effectiveness of the proposed algorithm on three Case studies. In the first two, simulation models were constructed based on standard biological practice. In the third, past observational data were used to generate synthetic data with two different strategies, and interventional data were used to verify the validity of the results.

\section{Background}

\subsection{Graphical notation and causal inference}
\label{sec:GraphNotCausalInf}
\begin{figure}[t]
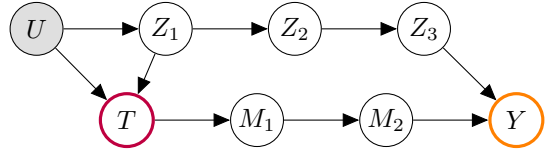

\begin{center}
\tikz[]{
     \node[latent] (Z1) {$Z_1$};%
     \node[obs, left=of Z1] (U) {$U$};
     \node[latent,right=of Z1] (Z2) {$Z_2$};%
     \node[latent,right=of Z2] (Z3) {$Z_3$};%
     \node[latent,below right=of Z1] (M1) {$M_1$};%
     \node[latent, right=of M1] (M2) {$M_2$};%
     \node[latent, draw = purple, very thick, left=of M1] (T) {$T$};%
     \node[latent, draw=orange, very thick, right=of M2] (Y) {$Y$};%
     \edge {Z1} {Z2};
     \edge {Z2} {Z3};
     \edge {M1} {M2};
     \edge {T}{M1};
     \edge {Z1}{T};
     \edge {Z3}{Y};
     \edge {M2}{Y};
     \edge{U}{T};
     \edge{U}{Z1};
   
     }
\end{center}
    \caption{\small  A directed acyclic graph (DAG)  with $3$ observable confounders ($Z_1$, $Z_2$, $Z_3$) and one latent confounder ($U$).
    White nodes are observable and grey nodes are latent.
    $T$ is the treatment and $Y$ is the effect.
    $M_1$ and $M_2$ are mediators.
    $Y$ is a collider on the colliding path $M_1 \rightarrow M_2 \rightarrow Y \leftarrow Z_3$.
    \label{fig:multConfounderMultMediator}}
\end{figure}

Let boldface letters such as $\mathbf{X}$ be a set of random variables and non-boldface letters such as $X$ be a random variable.
Let $x$ be an instance of  $X$,  and $\mathbf{x}$ an instance of  $\mathbf{X}$.
Let $G = (\mathbf{V} \cup \mathbf{U},\mathbf{E})$ be a directed acyclic graph (DAG) as in~\figref{fig:multConfounderMultMediator}, 
where $\mathbf{V}$ are observable variables (white nodes), and $\mathbf{U}$ are latent (grey nodes), and
$\mathbf{E}$ is a set of edges.
Let $P(\mathbf{V})$ be the joint distribution over the observable variables.

An \textbf{intervention} (perturbation) on a target variable in the graph (treatment) $T$ fixes it to a constant  value $t$ (denoted $do(T = t)$), and makes it independent of its causes~\citep{spirtes2000causation, eberhardt2007interventions}. 

A \textbf{causal query} $Q_{G}$ over the effect $Y$ is any probabilistic query that conditions on an intervention, such as  $Q_{G}=E[Y | do(T=t)]$ or $E[Y| do(T=1)] - E[Y| do(T=0)]$. 
The latter is a special case of causal query for binary treatments called \textbf{average treatment effect (ATE)} \cite{imbens2015causal}. 
Such causal queries are the focus of this manuscript.
A causal query $Q_{G}$ is \textbf{identifiable} from $P(\mathbf{V})$ compatible with a causal graph $G$ if the query can be computed uniquely \cite{pearl2009causality}. 

A \textbf{path} between two variables $T$ and $Y$ exists if there is a sequence of edges connecting $T$ to $Y$. 
A \textbf{directed path} follows the direction of the edges.
A \textbf{causal path} from $T$ to $Y$ is a directed path from $T$ to $Y$.
Let $cp(T, Y, G)$ be all the variables on causal paths from $T$ to $Y$ in $G$ excluding $T$. 
E.g. in \figref{fig:multConfounderMultMediator}, $cp(T,Y,G) = \{M_1, M_2, Y\}$. 
The \textbf{mediators} are all the variables on $cp(T, Y, G)$ except $T$ and $Y$.


{\bf d-separation} \cite{pearl2009causality} captures true independencies (i.e., separations) in a path.
A path $p$ is d-separated (or blocked) by a set of variables $\mathbf{Z}$ if and only if,
\vspace{-3mm}
\begin{itemize}
    \item $p$ contains a chain $V_i \rightarrow V_j \rightarrow V_k$ or a fork $V_i \leftarrow V_j \rightarrow V_k$ such that the middle variable
    $V_j$ is in $\mathbf{Z}$, or
    \vspace{-2mm}
    \item $p$ contains a collider $V_i \rightarrow V_j \leftarrow V_k$ such that the middle variable
    $V_j$ is not in $\mathbf{Z}$ and such that no descendant of $V_j$ is in $\mathbf{Z}$.
\end{itemize}
\vspace{-3mm}
A set $\mathbf{Z}$ d-separates $T$ from $Y$ if and only if $\mathbf{Z}$ blocks every path from $T$ to $Y$. 
For example in \figref{fig:multConfounderMultMediator}, any subset of $\{Z_1, Z_2, Z_3\}$ blocks the path $T \leftarrow Z_1 \rightarrow Z_2 \rightarrow Z_3 \rightarrow Y$.

In order to quantify the direct effect of $T$ on $Y$, one must isolate the effect of any possible 
confounders. 
A \textbf{confounder} is a variable that affects both $T$ and $Y$. 
E.g., in \figref{fig:multConfounderMultMediator}, $Z_1$, $Z_2$, and $Z_3$ are confounders. 
A path with arrows coming into both $T$ and $Y$ is called a \textbf{backdoor path}. In \figref{fig:multConfounderMultMediator} $T \leftarrow Z_1 \rightarrow Z_2 \rightarrow Z_3 \rightarrow Y$ is a backdoor path. 
The easiest way to eliminate confounding is to block all backdoor paths.

\begin{figure}[t]
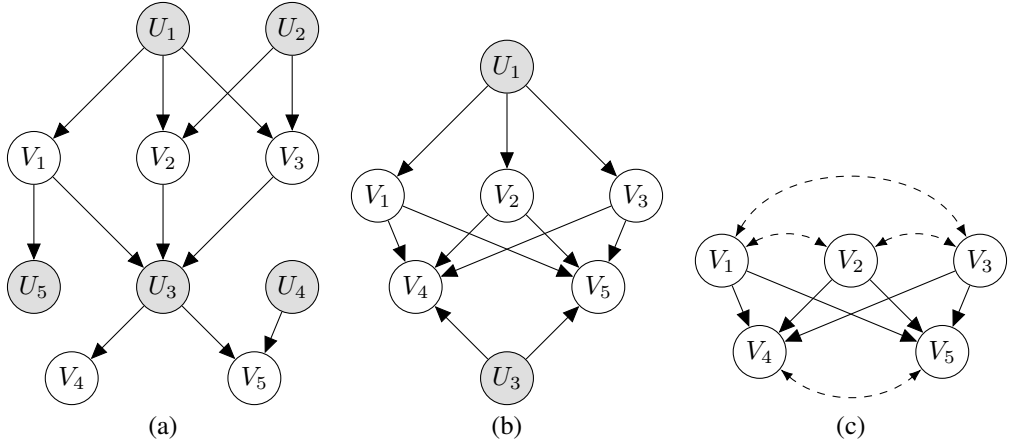

\begin{center}
\begin{tabular}{ccc}
\tikz[]{
     \node[obs] (U1) {$U_1$};%
     \node[obs,right=of U1] (U2) {$U_2$};%
     \node[latent,below=of U1] (V2) {$V_2$};%
     \node[latent, left=of V2] (V1) {$V_1$};%
     \node[latent, right=of V2] (V3) {$V_3$};%
     \node[obs,below=of V2] (U3) {$U_3$};%
     \node[latent,below left=of U3] (V4) {$V_4$};%
     \node[latent,below right=of U3] (V5) {$V_5$};%
     \node[obs,right=of U3] (U4) {$U_4$};%
     \node[obs,below=of V1] (U5) {$U_5$};%
     \edge {U1} {V1};
     \edge {U1} {V2};
     \edge {U1} {V3};
     \edge {U2} {V2};
     \edge {U2}{V3};
     \edge {V1}{U3};
     \edge {V2}{U3};
     \edge {V3}{U3};
     \edge {U3}{V4};
     \edge {U3}{V5};
     \edge {U4}{V5};
     \edge {V1}{U5};
     } &
\tikz[]{
     \node[obs] (U1) {$U_1$};%
     \node[latent,below=of U1] (V2) {$V_2$};%
     \node[latent, left=of V2] (V1) {$V_1$};%
     \node[latent, right=of V2] (V3) {$V_3$};%
     \node[latent,below left=of V2] (V4) {$V_4$};%
     \node[latent,below right=of V2] (V5) {$V_5$};%
      \node[obs,below right=of V4] (U3) {$U_3$};%
     \edge {U1} {V1};
     \edge {U1} {V2};
     \edge {U1} {V3};
     \edge {V1}{V4};
     \edge {V1}{V5};
     \edge {V2}{V4};
     \edge {V2}{V5};
     \edge {V3}{V4};
     \edge {V3}{V5};
     \edge {U3}{V4};
     \edge {U3}{V5};
     } &
\tikz[]{
     \node[latent,below=of U1] (V2) {$V_2$};%
     \node[latent, left=of V2] (V1) {$V_1$};%
     \node[latent, right=of V2] (V3) {$V_3$};%
     \node[latent,below left=of V2] (V4) {$V_4$};%
     \node[latent,below right=of V2] (V5) {$V_5$};%
     \edge {V1}{V4};
     \edge {V1}{V5};
     \edge {V2}{V4};
     \edge {V2}{V5};
     \edge {V3}{V4};
     \edge {V3}{V5};
    \path[bidirected] (V1) edge[bend left=30] (V2);
    \path[bidirected] (V2) edge[bend left=30] (V3);
    \path[bidirected] (V1) edge[bend left=60] (V3);
    \path[bidirected] (V4) edge[bend right=40] (V5);
     }\\
(a) & (b) & (c)
\end{tabular}
\end{center}
    \caption{\small (a) DAG $G$. White nodes are observable and grey nodes are unobserved (latent).
                    (b) Simplified sub-DAG $\widetilde{G}$ according to simplification rules.
                    (c) Corresponding ADMG $\mathcal{G}$. Bi-directed edges show presence of latent variable. The exact number and structure of latent variables in an ADMG can be misspecified.
    \label{fig:SimplifiedRulesAndADMG}}
\end{figure}

Frequently, some variables in a DAG are unobserved (i.e., latent) such as $U$ in~\figref{fig:SimplifiedRulesAndADMG}(a). DAGs with latent variables are compactly represented by \textbf{acyclic directed mixed graphs} (ADMGs)~\cite{richardson_2017} such as in~\figref{fig:SimplifiedRulesAndADMG}(c). An ADMG $\mathcal{G} = (\mathbf{V},\mathbf{E}_d, \mathbf{E}_b)$ consists of a set of observable variables $\mathbf{V}$, a set of directed edges $\mathbf{E}_d$, and bidirected edges $\mathbf{E}_b$. 
A bidirected edge indicates a path  including any number of latent variables without colliders that point to two observable variables.  
ADMGs allow us to misspecify the number and type of the latent variables, as long as we accurately represent the topology over the observed variables. ADMG is the main graph structure in this manuscript.

An ADMG is constructed by using the following \textbf{simplification rules} \cite{evans2016graphs}:  

\textbf{1.} Remove latent variables with no children from the graph.
E.g., $U_5$ in \figref{fig:SimplifiedRulesAndADMG}(a) is removed in \figref{fig:SimplifiedRulesAndADMG}(b).

\textbf{2.} Transform a latent variable with parents to an exogenous variable where all its parents are connected to its children. 
E.g., $U_3$ in \figref{fig:SimplifiedRulesAndADMG}(a) and (b).

\textbf{3.} Remove an exogenous latent variable that has at most one child. 
E.g., $U_4$ in \figref{fig:SimplifiedRulesAndADMG}(a) is removed in \figref{fig:SimplifiedRulesAndADMG}(b).

\textbf{4.} If $U$, $W$ are latent variables where children of $W$ are a subset of children of $U$, then, $W$ can be removed.
E.g.,
$U_1$ and $U_2$ in \figref{fig:SimplifiedRulesAndADMG}(a), and
$U_2$ is removed in \figref{fig:SimplifiedRulesAndADMG}(b).

\subsection{Input variables for causal query estimators}
\label{sec:CausalQueryEstProperties}

Given an ADMG and a causal query of interest, our objective is to determine which variables in the DAG should be used as input to a causal query estimator.
The choice of the variables determines the estimator's bias and variance.
An estimator is \textbf{biased} if it systematically deviates from its true value.
\textbf{Variance} of the estimator is its variability across repeatedly collected datasets.

The set of variables blocking the backdoor paths is called a \textbf{valid adjustment set} (e.g., $\{Z_1, Z_2, Z_3\}$ in \figref{fig:multConfounderMultMediator}).
The smallest set of variables that block all the back-door paths is called a \textbf{minimal adjustment set} (e.g., $Z_1$ or $Z_2$ or $Z_3$ in \figref{fig:multConfounderMultMediator}).
However, different adjustment sets produce estimators with different variance.

A minimal subset of the variables that blocks all back-door paths and maximally reduces the variability in $Y$ is an \textbf{optimal adjustment set} (e.g., $Z_3$ in \figref{fig:multConfounderMultMediator}). In absence of latent variables, the optimal adjustment set $ O(T,Y, G)$ is defined as \citep{rotnitzky2020efficient}
\begin{eqnarray}
    O(T,Y, G) = Pa(cp(T,Y,G)) \backslash (de(cp(T,Y,G)) \cup T) \label{eq:optAdjSet}
\end{eqnarray}
Here $Pa(cp(T, Y, G))$ are the parents, and $de(cp(T, Y, G))$ are the descendants of all the variables in $cp(T, Y, G)$. A valid, minimal, or optimal adjustment set can be computed with open-source R packages such as \texttt{pcalg} \cite{kalisch2012causal}, \texttt{dagitty} \cite{textor2016robust}, and \texttt{causaleffect} \cite{causal2017effect}.

In presence of latent variables,
\cite{runge2021necessary} showed that an optimal adjustment set exist in the specific case of linear models and under specific conditions. Beyond that, an optimal adjustment set does not exist in most scenarios. However, \cite{cinelli2020crash} provided graphical guidelines for determining adjustment sets that do not cause bias and possibly reduce or increase variance, as follows.

\textbf{Bad controls}, when added to the adjustment set, increase the bias of the query estimators. 
These variables
(1) open a backdoor path between $T$ and $Y$ (i.e, are colliders),
(2) are mediators or children of the mediators,
(3) open a path that contains a collider between $T$ and $Y$,
(4) are in $de(Y)$.

\textbf{Neutral controls}, when added to the adjustment set, neither increase nor decrease the bias, but may either increase or decrease the variance of the estimators.
\textbf{Good neutral controls} are possibly good for precision. These are variables that
(1) when added to the adjustment set, do not undermine the identifiability of the query, and (2) are in $O(T,Y, G)$. 
\textbf{Bad neutral controls} are possibly bad for precision. These are variables that
1) not necessary for the identifiability of the query, and 2) are in $Pa_{G}(T)$~\cite{hahn2004functional, white2011causal, henckel2019graphical}.

\subsection{Existing causal query estimators}
\label{sec:queryEstimationApproaches}

Given an ADMG, and a causal query of interest, the next objective is to choose an estimator for the query.
Many estimators are implemented in open-source libraries such as {\texttt{DoWhy}}~\cite{dowhy},
{\texttt{Ananke}}~\cite{bhattacharya2020semiparametric},
and the \href{https://y0.readthedocs.io}{\texttt{$Y_0$}  engine}~\cite{y0}. 
In this manuscript we focus on the following.

{\bf Non-parametric and semi-parametric estimators} such as gformula, inverse probability weight (IPW), augmented IPW (AIPW), Nested IPW and augmented nested IPW~\citep{bhattacharya2020semiparametric} are all implemented and well-documented in {\texttt{Ananke}}.
These estimators are asymptotically unbiased, do not require parametric assumptions, and are computationally cheap.
However, they are limited to causal queries with one treatment and one effect, and the treatment must be binary-valued.
Moreover, they limit the variables used for query estimation. 
For example, the IPW estimator only uses the cause, effect and a valid adjustment set. The rest of variables such as mediators and the variables not in the adjustment set are ignored.


{\bf Causal generative models} expand causal query estimation beyond a single binary treatment and a single effect. \cite{mohammad2022calculus} represented the data generating process with a directed graphical model. 
The approach 1) estimates the posterior distribution over the model parameters given the training data, 
2) fixes the targets of the intervention, and breaks their relationship to their parents, and
3) samples the parameters from their posterior distributions, and then samples from each variable given its parents. 
The estimator can be thought of as a posterior predictive statistic over the marginal of the parameters.
The query is estimated in an asymptotically unbiased manner with all the variables, except for the ancestors of the treatment $T$ that do not affect the descendants of $T$, and except for the descendants of the effect. 
Unfortunately, the approach is computationally expensive and requires parametric assumptions.

\textbf{Linear regression without mediators} simplifies the complexity of causal generative models by only regressing $T$ and adjustment set on $Y$.
The coefficient of $T$ is an unbiased estimator of the ATE.
For example in~\figref{fig:multConfounderMultMediator},
\begin{eqnarray}
\label{eq:simpleLinearEq}
    Y = \beta_0 + \beta T + \gamma Z_i + \epsilon; \epsilon \stackrel{iid}{\sim} \mathcal{N}(0,\sigma^2) \label{eq:lmWithMed}
\end{eqnarray}
The least squares estimator $\hat{\beta}$ estimates the ATE of $T$ on $Y$ without bias. 
Even after adding more than one $Z_i$, $i \in \{1,2,3\}$ the estimator of the ATE remains unbiased because the backdoor path remains blocked. However its variance will change. 
$E[Y|do(T=t)]$ is estimated by substituting $\hat{\beta}$ and $\hat{\gamma}$ in \eqref{eq:lmWithMed} and averaging over the values of $Y$.

The linear regression without mediators approach is  simple, fast, and highly interpretable. 
It takes as input treatment, effect and any variables in the adjustment set. However, it is only appropriate when the variables are linearly related.

\textbf{Linear regression with mediators} allows for inclusion of mediators absent from an adjustment set.
Similarly to linear regression without mediators, it estimates the ATE of $V_i$ on $V_j$ for any pair of variables on the causal paths between $T$ and $Y$,
and multiply all the estimated ATEs.
In~\figref{fig:multConfounderMultMediator},
\begin{eqnarray}
    M_1 &=&  \alpha_0 +\alpha T + \nu;\ \nu \stackrel{iid}{\sim} \mathcal{N}(0,\sigma^2_\nu)  \label{eq:IncludMed1}\\
    M_2 &=&  \rho_0 + \rho M_{1} + \psi;\ \psi \stackrel{iid}{\sim} \mathcal{N}(0,\sigma^2_\psi)  \label{eq:IncludMed2}\\
    Y   &=&  \eta_0 + \eta M_2 + \delta Z_i + \omega;\ \omega \stackrel{iid}{\sim} \mathcal{N}(0,\sigma^2_\omega) \label{eq:IncludMed3}
\end{eqnarray}
By substituting \eqref{eq:IncludMed1} into \eqref{eq:IncludMed2}, and \eqref{eq:IncludMed2} into \eqref{eq:IncludMed3}
\begin{align}
    Y   &= \eta_0 + \eta \rho_0 + \eta \rho \alpha_0 +(\eta \rho \alpha) T + \delta Z_i + \eta \rho \nu + \eta \psi + \omega
    = \zeta_0 + \zeta T + \delta Z_i + \epsilon' \label{eq:IncludMed4}
\end{align}
where $\ \epsilon' \stackrel{iid}{\sim} \mathcal{N}(0,\eta^2 \rho^2 \sigma^2_\nu + \eta^2 \sigma^2_\psi + \sigma^2_\omega)$. 
We establish that $\eta \rho \alpha$ is an ATE, and that the least squares estimator of $\eta \rho \alpha$ is unbiased. Estimating $\eta$, $\rho$ and $\alpha$ separately and multiplying the estimates reduces the variance of the estimator \cite{bellemare2019paper}.

\begin{figure*}[t!]
\begin{center}
\begin{tabular}{c}
\includegraphics[scale=0.4]{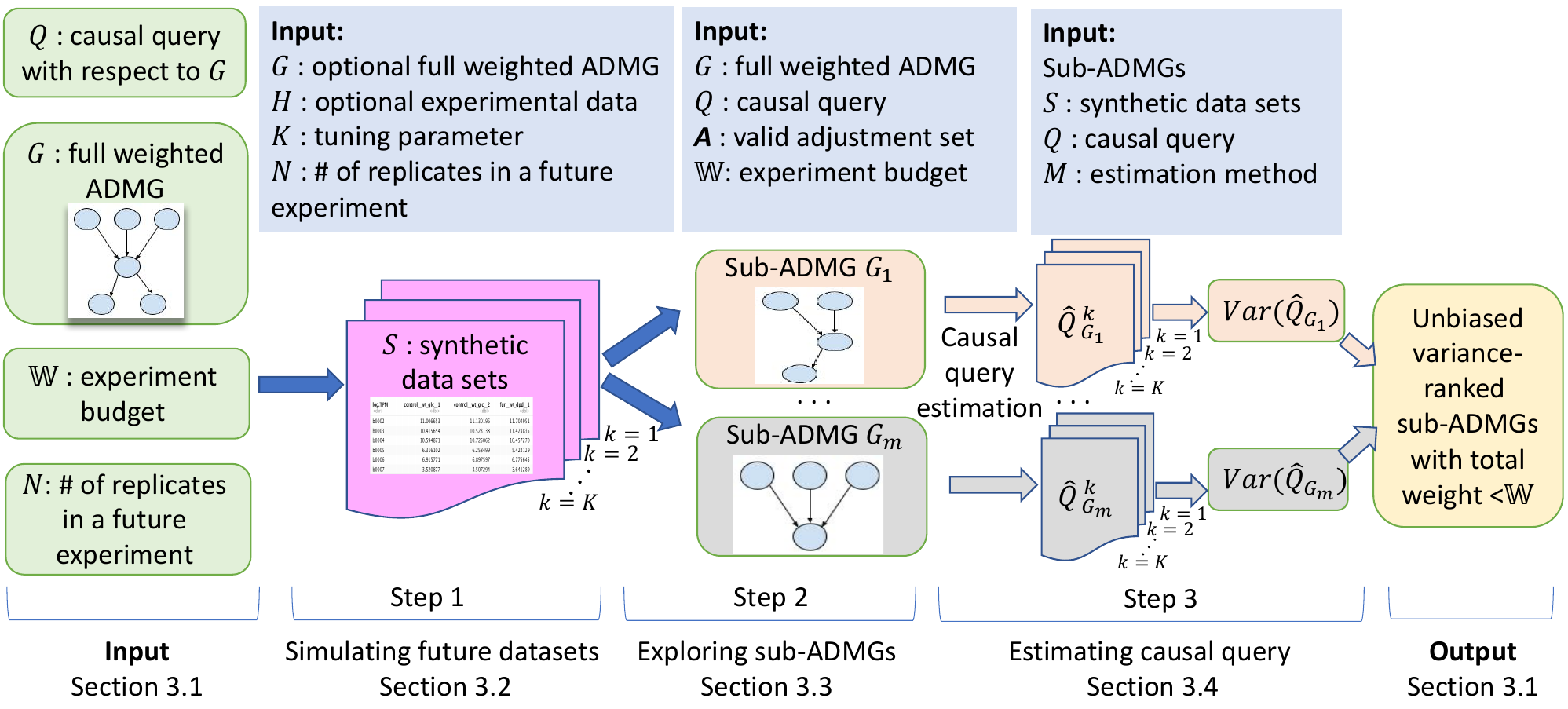}
\end{tabular}
\end{center}
    \caption{\small  
    Overview of the proposed approach. Arrows indicate sequence of steps.
    Inputs specific to the biological problem are shown in green. Additional tuning parameters and analytical options specified for each step are shown in blue.
    \label{fig:overview}
    }
\end{figure*}

\subsection{Simulation-based inference}

Synthetic data representative of a particular experiment is an increasingly popular strategy for experiment planning and analysis. 
Generative models such as GANs~\cite{goodfellow2020generative} successfully generated high-quality artificial genomes~\cite{yelmen2019creating}, MRI scans~\cite{pawlowski2020deep}, and electronic health records while mitigating privacy concerns~\cite{choi2017generating,weldon2021generation}.

Numerous single-cell RNA (scRNA-seq) sequencing data simulation methods~\cite{marouf2020realistic} have been developed to generate realistic scRNA sequencing data, and  to produce data with desired structure (such as specific cell types or subpopulations) while preserving genes, capturing gene correlations, and generating any number of cells with varying sequencing depths~\cite{cao2021benchmark,sun2021scdesign2}. They were also useful to augment sparse cell populations and improve the quality and robustness of downstream classification. To the best of our knowledge, there are currently no approaches for simulation-based design of experiments for causal query estimation.



\section{Methods}
\label{sec:methods}

We propose an approach that assists with planning a future experiment for causal query estimation. 
We combine the existing methods for finding optimal adjustment sets, and synthetic data generation. 
We report sub-ADMGs supporting (asymptotically) unbiased estimation of a causal query, and ranked according to variance of the causal query estimator and experimental cost. 
The user can select for future measurement a sub-ADMG that balances variance of the query and cost. 
\figref{fig:overview} overviews the approach. 

\subsection{Input and output}
The proposed approach takes as input a 
weighted ADMG $G$ from a knowledge base such as INDRA or Omnipath. 
The ADMG can misspecify the structure of the latent variables, as long as it accurately represents the structure over the observable variables.
The observable variables of the ADMG are weighted by their associated experimental cost.

The proposed approach also takes as input a causal query $Q$ with respect to $G$ of the form $E[Y|do(T=t)]$ or $ATE = E[Y|do(T=t+1)] - E[Y|do(T=t)]$ for a given $t$,
the upper limit on the experimental budget $\mathbb{W}$, and the number of replicates $N$ in the experiment being planned.

The proposed approach outputs a list of sub-ADMGs $G_1, ..., G_m$ that support asymptotically unbiased estimation of the query and satisfy the budget constraint. 
The sub-ADMGs are ranked by the empirically evaluated variance of the causal query estimators. 

\subsection{Step 1 : Simulating future data sets}

The approaches in \secref{sec:CausalQueryEstProperties} that determine input variables for causal query estimation have limitations. 
They focus on graphical criteria, and do not consider variables such as mediators that can't be part of an adjustment set.
Furthermore, in presence of latent variables no uniformly optimal adjustment set may exist~\cite{henckel2019graphical, rotnitzky2020efficient}.
Therefore, we evaluate the estimators by generating $K$ synthetic datasets that mimic the future experiment, with $N$ observations each.
The methodological challenges are the choice of the simulator, and the diagnostics of the simulation accuracy. 
$K$ is a tuning parameter.

When historical experimental data (e.g., prior data from the same organism, or from related systems or technologies) exist,
the synthetic data is constructed based on approaches such as Tabular GAN \cite{xu2019modeling}, or based on statistical associations such as Bayesian network forward simulation~\cite{koller2009probabilistic}.
The latter requires as input the knowledge of the ADMG.
The quality of the synthetic data is evaluated by comparing the marginal and the joint distributions of the synthetic and the experimental data.
In absence of historical data, the synthetic data sets are constructed using reasonable or known ranges for model parameters, and functional forms such as Hill equations \cite{alon2019introduction}. 
The quality is evaluated by sensitivity analysis across multiple values of parameters and functional forms.


\subsection{Step 2 : Exploring sub-ADMGs}

{\bf Description of the algorithm} This step generates sub-ADMGs that support (asymptotically) unbiased estimation of the causal query and do not exceed the budget. The biological inputs to the step are the full weighted ADMG, the causal query of interest,  and the experimental budget.
The technical input is a valid adjustment set that includes all the good neutral controls that are proven to possibly decrease the variance of the estimation. 
\algref{alg:generateValidAdjustmentSet} in Appendix describes the generation of the adjustment set. 
We call the treatment, effect, and the adjustment set the \textbf{required variables}, necessary for an (asymptotically) unbiased estimation of the query. 
These variables are always present in the search space. 
The rest of (yet unexplored) variables are \textbf{optional nodes}.

The remainder of Step 2 is detailed in \algref{alg:generateSetOfRankedSubADMGs}.
Since unidentifiable queries provide biased estimators, the algorithm raises an error if the input query is unidentifiable (\lineref{line:IfNotIdentifiable}).
Descendants of $Y$ (i.e., bad controls in \secref{sec:CausalQueryEstProperties}) can't cause any variation in $Y$, and also bias the results.
Hence, we only consider the ancestral graph of $Y$ (\lineref{line:ancestralOfY}).

The algorithm iterates over subsets of nodes in the ancestral graph limited to those which contain all the required nodes and have sum of weights less than the given budget (\lineref{forS}, \algref{alg:generateValidSubsets} and \algref{alg:generateValidSubsetsHelper} in Appendix).
For each of these subsets a sub-ADMG is generated (\lineref{genSUbADMG}, \algref{alg:generateSubADMG} in Appendix).

{\bf Properties of the algorithm}
\algref{alg:generateSetOfRankedSubADMGs} enumerates all the sub-ADMGs satisfying the budget constraint, and finds the sub-ADMG(s) with globally minimal variance of the causal query estimator. 
Lemma 1 in Appendix shows that including any number of mediators as input to the causal query estimator always reduces the variance.
Therefore, the optimal sub-ADMG always includes mediators.

\algref{alg:generateSetOfRankedSubADMGs} is of exponential complexity. 
For $\mathbb{W}=\infty$, the running time complexity is of the order of $O(n\times2^{n})$, where $n$ is the total number of optional variables.
Setting $\mathbb{W}<\infty$ limits the search space and reduces the complexity.

Despite the exponential complexity, \algref{alg:generateSetOfRankedSubADMGs} supports many practically relevant queries. The case studies in this manuscript took between 2 to 10 minutes on a parallel Google cloud Platform with 2 vCPUs and 8 GB memory.

For large-scale problems, we can take additional shortcuts.
As we show in the Case studies, a single mediator substantially decreases the variance of the causal query estimator.
For every additional mediator the decrease is smaller.
Therefore, searching over sub-ADMGs with a single mediator can be effective.
Another shortcut is to stop the algorithm once the selected variables reduce the variance with respect to the required variables  by a pre-specified amount.
These shortcuts do not produce globally optimal sub-ADMGs, but reduce the experimental and computational cost.


\begin{footnotesize}
\begin{algorithm}[ht!]
\small
\caption{generateSetOfRankedSubADMGs}
\label{alg:generateSetOfRankedSubADMGs}
\DontPrintSemicolon
\SetAlgoLined
\KwInput{
         $\mathcal{G} = (\mathbf{V}, \mathbf{E}_d,  \mathbf{E}_b, \mathbf{W})$: Weighted ADMG\\
         $~~~~~~~~~~~~~$ $T$,\  $Y$: Treatment and effect variables \\
         $~~~~~~~~~~~~~$ $Q_{\mathcal{G}}$: Causal query \\
         $~~~~~~~~~~~~~$ $\mathbf{A}$: A valid adjustment set $~~${\color{brown} \ttfamily \scriptsize //Output of Alg7 in Appx}\\ 
         $~~~~~~~~~~~~~$ $\mathbf{D}$: Input data sets {\color{brown} \ttfamily \scriptsize // Step 1}\\
         $~~~~~~~~~~~~~$ $M$: Causal query estimator 
}
\KwParameter{ $\mathbb{W}$: Upper limit on the total weight }

\KwOutput{A set of variance-ranked sub-ADMGs with total weight $\le$ $\mathbb{W}$. $T$, $Y$ and $\mathbf{A}$  are always present.}

\vspace{0.2cm}
\hrule 
\vspace{0.2cm}
\setstretch{1.2} 

\If{$Q_{\mathcal{G}}$ is not identifiable \label{line:IfNotIdentifiable}}{\textbf{raise} not identifiable error} \label{line:EndIfNotIdentifiable}

$\mathcal{G} \leftarrow \mathcal{G}_{an{(Y)}}$ {\color{brown} \ttfamily \scriptsize // Ancestral graph of $Y$} \label{line:ancestralOfY}

$\mathbf{subADMGSet} \leftarrow$  \{\}
        
 \For{$\mathbf{s} \in$ \normalfont{getValidNodeSubsets}( $\mathcal{G}, \mathbf{A}, T, Y,\mathbb{W}$) \label{forS}}{

    $\mathcal{G'}$ = generateSubADMG($\mathcal{G}$, $\mathbf{s}$) \label{genSUbADMG}  {\color{brown} \ttfamily \scriptsize //Step 2}
    
     
    $\widehat{Q}_{\mathcal{G}}$ = estimateQuery($\mathcal{G'}$, $Q_{\mathcal{G}}$, $\mathbf{D}$, $M$) \label{estimateQuery} {\color{brown} \ttfamily \scriptsize //Step 3}
            
    $\mathbf{subADMGSet}.add\{ \mathcal{G'}:var(\widehat{Q}_{\mathcal{G}}) \}$ \label{addADMGVar}
   
    }
\textbf{return} sorted $\mathbf{subADMGSet}$ based on variance
\label{line:returnGenerateSetOfRankedSubADMGs}
        
\end{algorithm}
\end{footnotesize}

\subsection{Step 3 : Estimating causal query}

This step estimates the query for each sub-ADMG over all the $K$ synthetic data sets (\algref{alg:generateSetOfRankedSubADMGs}, \lineref{estimateQuery}). 
The technical input is the choice of a query estimator.
Here we consider all the estimators in \secref{sec:queryEstimationApproaches}.
For the semi- and non-parametric approaches, \algref{alg:generateSetOfRankedSubADMGs} only searches over the variables that these estimators use and disregard the rest.
Finally, we calculate the empirical variance for each sub-ADMG (\algref{alg:generateSetOfRankedSubADMGs}, \lineref{addADMGVar}).
When \algref{alg:generateSetOfRankedSubADMGs} repeated with different causal query estimators and different number of replicates $N$, we can evaluate the impact of the sample size and of the estimator.


\subsection{Illustrative example} 
\label{sec:IllustrativeExample}

This simple example illustrates the impact of using a subset of variables to measure on a causal query estimation, in an idealized case where the data generation process is known.

\begin{figure}[t]
\begin{center}
\begin{tabular}{cc}
\tikz[]{
     \node[latent] (Z1) at (0,0) {$Z_1$};%
     \node[latent] (Z2) at (1.5,0) {$Z_2$};%
     \node[latent] (Z3) at (3,0) {$Z_3$};%
     \node[latent] (Z4) at (4.5,0){$Z_4$};%
     \node[latent] (Z5) at (6,0) {$Z_5$};%
     \node[latent,draw=purple,very thick] at (0,-1.5) (T) {$T$};%
     \node[latent] (M1) at (1.5,-1.5) {$M_1$};%
     \node[latent] (M2) at (3,-1.5) {$M_2$};%
     \node[latent] (Z6) at (3,-3) {$Z_6$};%
     \node[latent] (M3) at (4.5,-1.5) {$M_3$};%
     \node[latent] (Z7) at (4.5,-3) {$Z_7$};%
     \node[latent,draw=orange,very thick] (Y)  at (6,-1.5) {$Y$};%
     \edge {Z1} {Z2};
     \edge {Z2} {Z3};
     \edge {Z3} {Z4};
     \edge {Z4} {Z5};
     \edge {T} {M1};
     \edge {M1} {M2};
     \edge {M2} {M3};
     \edge {M3} {Y};
     \edge {M3} {Z7};
     \edge {Z1}{T};
     \edge {Z2}{M1};
     \edge {Z5}{Y};
     \edge {T}{Z6};
     \edge {Y}{Z6};
     
    \path[bidirected] (Z1) edge[bend left=60] (T);
    \path[bidirected] (Z2) edge[bend left=60] (M1);
     }
     &

\includegraphics[scale=0.23]{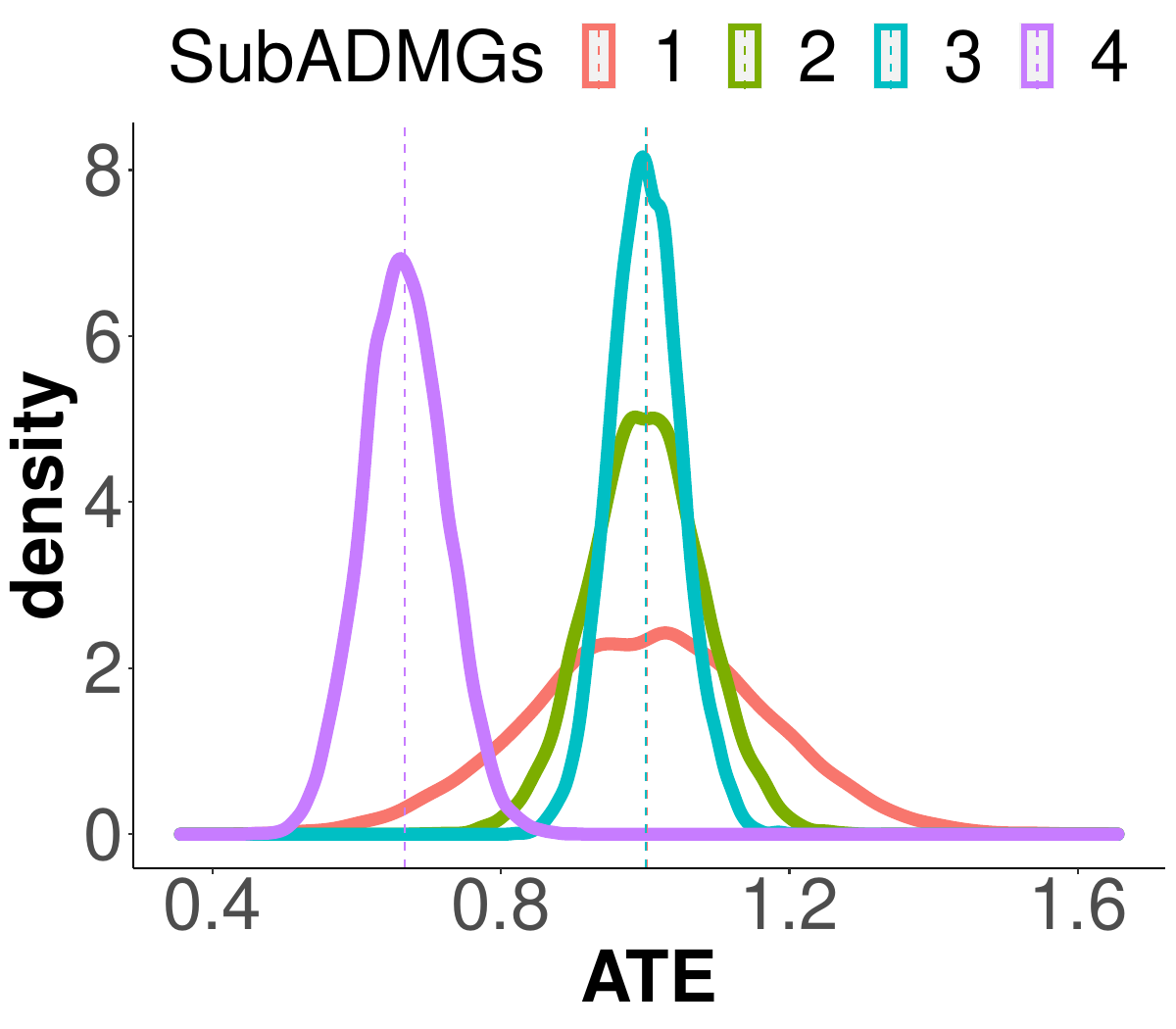} 
\\
(a) & (b) 
\end{tabular}
\end{center}
    \caption{\small \textbf{Illustrative example } 
    (a) The full ADMG, with treatment $T$ and effect $Y$.
    The minimal adjustment set is $Z_1$ and the optimal adjustment set is $\{Z_1, Z_2, Z_5\}$.
    (b) ATE density for linear regression estimates on over 10000 data sets for a selection of sub-ADMGs in \secref{sec:IllustrativeExample}, Step 2.
    True ATE=1.
    The optimal {\bf Sub-ADMG 3} consists of $T$, $Y$, $Z_1$, $Z_2$, $Z_5$, $M_1$, $M_2$.
    Estimate of ATE with the full ADMG (\textbf{Sub-ADMG 4}) biased the results.
    \label{fig:MotivExample1}
    }
\end{figure}

{\bf Ground truth} Consider the weighted ADMG in \figref{fig:MotivExample1}, where
$T$ is the treatment and $Y$ is the effect. 
The causal query of interest is the ATE.
Denote $\mathbf{X}$ the set of all the variables. Assume that the data generation process follows,
\begin{eqnarray}
\label{eq:mot1DataGen}
    X = \theta Pa(X) + N_X; N_X \stackrel{iid}{\sim} \mathcal{N}(0, 1)
\end{eqnarray}
with $\theta=1$.
Since the ATE is the coefficient of $T$, when $T$ and the adjustment sets are regressed on $Y$, the true ATE=1.

{\bf Input} Assume that the proposed approach takes as input the correct ADMG in \figref{fig:MotivExample1}, $T$, $Y$, and the query ATE.
Assume  that all the variables have the same cost of 1 (i.e., the total cost is the number of variables),  $\mathbb{W}=\infty$, and $N=500$.

\textbf{Step 1: Simulating future data sets} We generated 10000 synthetic data sets with 500 data points  from~\eqref{eq:mot1DataGen}.

\textbf{Step 2: Exploring sub-ADMGs} 
In this example the minimal valid adjustment set is $Z_1$, i.e. $Z_1$ is always present in the search space.
We hand-picked 4 sub-ADMGs that illustrate the proposed approach. They consist of,

1) \textbf{Sub-ADMG 1:}  $T$, $Y$, $Z_1$

2) \textbf{Sub-ADMG 2:} $T$, $Y$, $Z_1$, $Z_2$, $Z_5$

3) \textbf{Sub-ADMG 3:} $T$, $Y$, $Z_1$, $Z_2$, $Z_5$, $M_1$, $M_2$, $M_3$

4) \textbf{Sub-ADMG 4:} full ADMG in \figref{fig:MotivExample1} (a).

\textbf{Step 3: Estimating causal query}
For each simulated data set and each sub-ADMG, we estimated the ATE.
With the first sub-ADMG, the ATE is estimated by controlling for the effect of the confounders by blocking the backdoor path between $T$ and $Y$ by adjusting for $Z_1$,
\begin{eqnarray}
    Y = \beta T + \zeta Z_1 + \epsilon; \epsilon \stackrel{iid}{\sim} \mathcal{N}(0,1) \label{eq:YGivenTZ1}
\end{eqnarray}
Here $\widehat{ATE} = \hat{\beta}$ is an unbiased estimator .
With the second sub-ADMG, the ATE is estimated by controlling for the effect of the confounders by blocking the backdoor path with $Z_1$ while also including $Z_2$, $Z_5$.
\begin{eqnarray}
    Y &=& \delta T + \eta Z_1 + \kappa Z_2 + \phi Z_5 + \nu; \nu \stackrel{iid}{\sim} \mathcal{N}(0,1) \label{eq:YGivenTZ1Z2Z5}
\end{eqnarray}
Note that $Z_2$ and $Z_5$ are good neutral controls. 
In this case, $\widehat{ATE} = \hat{\delta}$ is unbiased and has smaller variance.


With the third sub-ADMG, the query is estimated by including the mediators ($M_1$, $M_2$, $M_3$):
\begin{eqnarray}
M_1 &=&  \alpha T  + \gamma Z_1 + \psi Z_2  + \omega; \omega \stackrel{iid}{\sim} \mathcal{N}(0,1)   \label{eq:M1GivenTZ1Z2*}\\
M_2 &=&  \digamma M_{1} + \varphi; \varphi \stackrel{iid}{\sim} \mathcal{N}(0,1)   \label{eq:M2GivenM1}\\
M_3 &=&  \rho M_{2} + \varrho; \varrho \stackrel{iid}{\sim} \mathcal{N}(0,1)   \label{eq:M3GivenM2}\\
Y   &=&  \chi M_3 + \varkappa Z_5 + \varepsilon; \varepsilon \stackrel{iid}{\sim} \mathcal{N}(0,1) \label{eq:YGivenM3Z5*}
\end{eqnarray}
Here $\widehat{ATE} = \hat{\alpha} \hat{\digamma} \hat{\rho} \hat{\chi}$ is unbiased estimator of ATE and the estimate has a smaller variance than the first two approach.
With the fourth sub-ADMG, the ATE is estimated by including all the variables into estimation of the query. 

{\bf Output and conclusions} \figref{fig:MotivExample1} (b) shows the results of query estimation over the synthetic data for each sub-ADMG. 
We can make several conclusions.
(1) As expected, the adjustment set that included the good neutral controls ($Z_2$, $Z_5$) reduced the variance compared to only measuring the minimal adjustment set ($Z_1$).
(2) Including the mediators reduced the variance beyond what was possible with the adjustment set.
Mediators are often viewed as bad controls, and not used for causal query estimation. 
This example illustrated that mediators can reduce the variance of estimators, at least in a linear setting.
(3) Using all the variables was detrimental. Since some of the included variables were bad controls ($Z_6$, $Z_7$), they biased the results.


\begin{figure*}[th]
\begin{center}
\begin{tabular}{llll}
\includegraphics[scale=0.4]{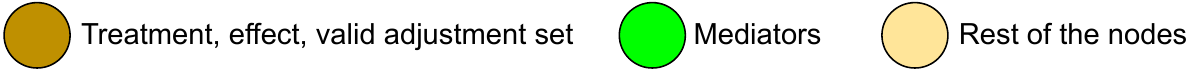} & {$~~~~~~~~~~~~~~~~~~~$} & {$~~~~~~~~~~~~~~~~~~~$} & {$~~~~~~~~~~~~~~~~~~~$}
\end{tabular}
\begin{tabular}{ccccc}
\resizebox{2.5 cm}{!}{%
\tikz[]{
     \node[fill=yellow,circle,draw,minimum size=1.1cm, label={[align=center,font=\Large]above left:30}] (SOS) {$\mathbf{SOS}$};%
     \node[fill=yellow,circle,draw,below=of SOS,minimum size=1.1cm, label={[align=center,font=\Large]above left:100}] (Ras) {$\mathbf{Ras}$};%
     \node[fill=green,circle,draw,below=of Ras,minimum size=1.1cm, label={[align=center,font=\Large]above left:60}] (Raf) {$\mathbf{Raf}$};%
     \node[fill=olive,circle,draw=purple,very thick,circle,left=of Raf,minimum size=1.1cm, label={[align=center,font=\Large]above left:1}] (Akt) {$\mathbf{Akt}$};%
     \node[fill=green,draw,circle,below=of Akt,minimum size=1.1cm, label={[align=center,font=\Large]above left:20}, 
     label={[align=right,font=\Large]below:$~~~~~~$total}] (Mek) {$\mathbf{Mek}$};%
     \node[fill=olive,circle,draw=orange,very thick, right=of Mek,minimum size=1.1cm, label={[align=center,font=\Large]above left:1},
     label={[align=left,font=\Large]below: weight:213}] (Erk) {$\mathbf{Erk}$};%
     \node[fill=olive,draw,circle,left=of Ras,minimum size=1.1cm, label={[align=center,font=\Large]above left:1}] (PI3K) {$\mathbf{PI3K}$};%

     \edge {SOS} {Ras};
     \edge {Ras} {Raf};
     \edge {Raf} {Mek};
     \edge {Mek} {Erk};
     \edge {Ras} {PI3K};
     \edge {PI3K} {Akt};
     \edge {Akt} {Raf};
     \filldraw[black, very thick] (-0.8,-4.5) rectangle (-.5,-3.8);
     
    \path[bidirected] (SOS) edge[bend right=60] (PI3K);
     }} &
\resizebox{1.9 cm}{!}{%
\tikz[]{
     \node (PI3K) at (-2,-2) [fill=olive,draw,circle,minimum size=1.1cm, label={[align=center,font=\Large]above left:1}]  {$\mathbf{PI3K}$};%
     \node (Akt) at (-2,-4) [fill=olive,circle,draw=purple,very thick,circle,minimum size=1.1cm, label={[align=center,font=\Large]above left:1}]  {$\mathbf{Akt}$};%
     \node (Erk) at (-2,-6) [fill=olive,circle,draw=orange,very thick,minimum size=1.1cm, label={[align=center,font=\Large]left:1},label={[align=center,font=\Large]below:total weight : 3}]  {$\mathbf{Erk}$};%
     
     \edge {Akt} {Erk};
     \edge {PI3K} {Akt};
    \path[bidirected] (PI3K) edge[bend left=60] (Erk);
     }} &
\resizebox{1.9 cm}{!}{%
\tikz[]{
     \node (PI3K) at (-2,-2) [fill=olive,draw,circle,minimum size=1.1cm, label={[align=center,font=\Large]above left:1}]  {$\mathbf{PI3K}$};%
     \node (Akt) at (-2,-4) [fill=olive,circle,draw=purple,very thick,circle,minimum size=1.1cm, label={[align=center,font=\Large]above left:1}]  {$\mathbf{Akt}$};%
     \node (Mek) at (-2,-6) [fill=green,draw,circle,minimum size=1.1cm, label={[align=center,font=\Large]above left:20}]  {$\mathbf{Mek}$};%
     \node (Erk) at (-2,-8) [fill=olive,circle,draw=orange,very thick,minimum size=1.1cm, label={[align=center,font=\Large]above left:1},label={[align=center,font=\Large]below:total weight : 23}]  {$\mathbf{Erk}$};%
     \edge {Mek} {Erk};
     \edge {Akt} {Mek};
     \edge {PI3K} {Akt};
   \path[bidirected] (Erk) edge[bend right=30] (PI3K);
     }} &
\resizebox{2 cm}{!}{%
\tikz[]{
     \node (SOS) at (-2,0) [fill=yellow,circle,draw,minimum size=1.1cm, label={[align=center,font=\Large]above left:30}]  {$\mathbf{SOS}$};%
     \node (PI3K) at (-2,-2) [fill=olive,draw,circle,minimum size=1.1cm, label={[align=center,font=\Large]above left:1}]  {$\mathbf{PI3K}$};%
     \node (Akt) at (-2,-4) [fill=olive,circle,draw=purple,very thick,circle,minimum size=1.1cm, label={[align=center,font=\Large]above left:1}]  {$\mathbf{Akt}$};%
     \node (Erk) at (-2,-6) [fill=olive,circle,draw=orange,very thick,minimum size=1.1cm, label={[align=center,font=\Large]left:1}, label={[align=center,font=\Large,font=\Large]below:{total weight : 33}}]  {$\mathbf{Erk}$};%

     \path[] (SOS) edge[bend left=30] (Erk);
     \edge {SOS} {PI3K};
     \edge {Akt} {Erk};
     \edge {PI3K} {Akt};
     
    \path[bidirected] (SOS) edge[bend right=60] (PI3K);
     }} &
     \includegraphics[scale=0.19]{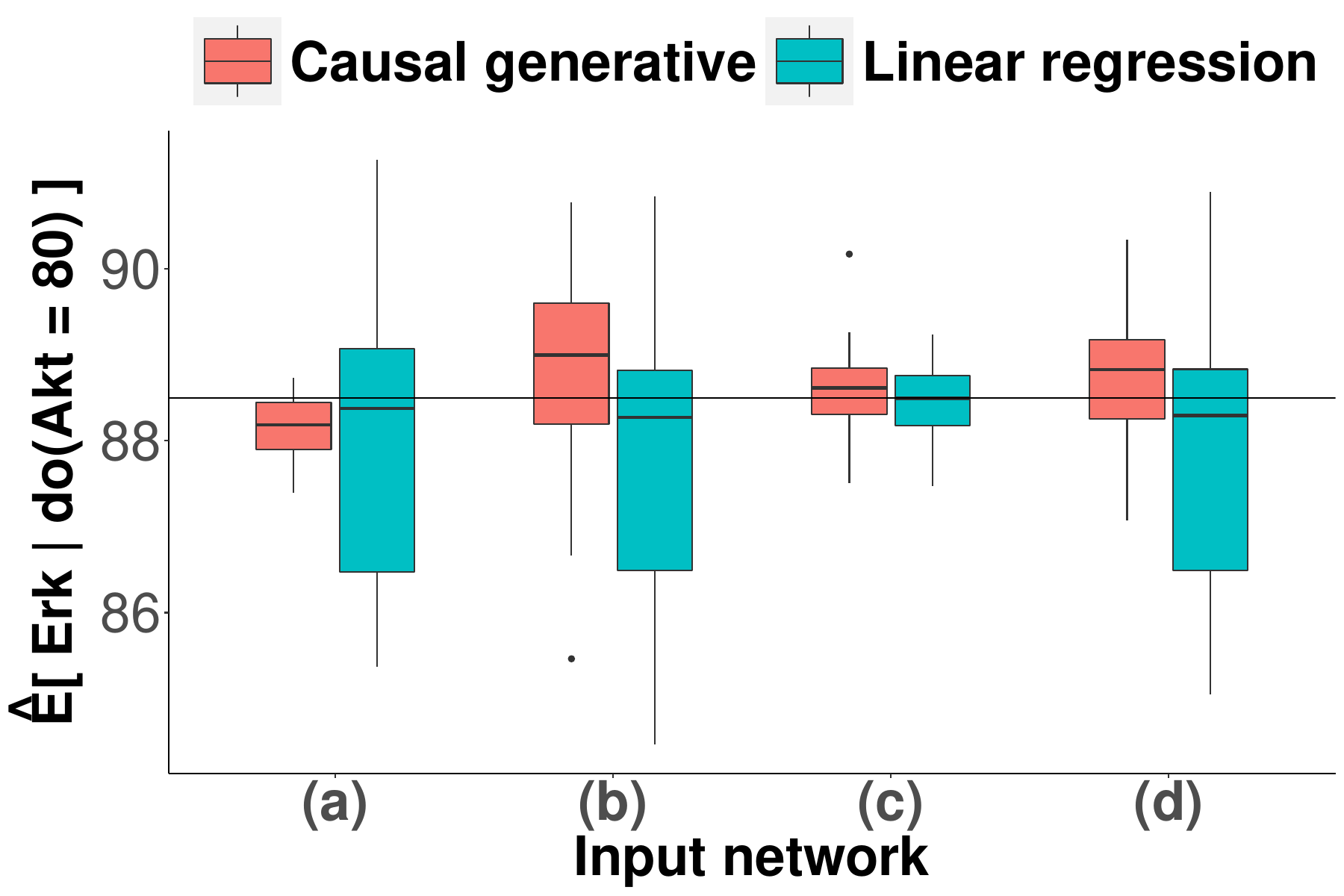}
     \\
 (a) & (b) & (c) & (d) & (e) 
\end{tabular}
\end{center}
    \caption{\small \textbf{Case study 1.} 
    (a) Full weighted ADMG. $Akt$ is the treatment and $Erk$ is the effect.
    Bi-directed edges indicate presence of latent variables.
    (b)-(d) Example sub-ADMGs.
    Simplification rules in \secref{sec:GraphNotCausalInf} lead \algref{alg:generateSetOfRankedSubADMGs} to convert the latent variables into exogenous variables, shown with bi-directed edges.
    (e) Sampling distribution of $\hat{E}[Erk | do(Akt=80)]$ over $K=20$ simulated observational data, $N=100$.
    The horizontal line is the true value of the query from interventional data.
    The apparent bias was due to the approximations in the model assumptions, and to the asymptotic nature of the estimators. 
    \label{fig:IGFgenerateSetOfRankedSubADMGs} }
\end{figure*}

\begin{table*}[th]
\centering
\begin{small}
\begin{tabular}{|c || c | c | c | c || c | c | c | c |} 
 \hline
& 
\multicolumn{4}{c||}{$\mathbf{N=20}$} &
\multicolumn{4}{c|}{$\mathbf{N=100}$}
 \\[1mm]
 \hline
 \textbf{Variance by} &
 \textbf{Fig. 5(a)} &
 \textbf{Fig. 5(b)} &
 \textbf{Fig. 5(c)} &
 \textbf{Fig. 5(d)} &
 \textbf{Fig. 5(a)} &
 \textbf{Fig. 5(b)} &
 \textbf{Fig. 5(c)} &
 \textbf{Fig. 5(d)} \\
 \hline\hline
 \textbf{Linear regression} &
 21.72 &  18.40 & \cellcolor{blue!25} 1.67 & 19.78 &   
 2.39 & 2.41 & \cellcolor{blue!25} 0.236 &  2.25  \\
 \hline
 \textbf{Causal generative} &
1.22 & 4.83 & \cellcolor{blue!25} 0.75 & 6.95  &   
\cellcolor{blue!25} 0.15 & 1.60 & 0.33 &  0.66 \\
\hline
\end{tabular}
\end{small}
\caption{\small \textbf{Case study 1.} Impact of the choice of the estimator, and of $N$. Highlighted cells show minimal variance.
\label{tables:IGF}}
\end{table*}

\section{Case studies of biomolecular networks}

We illustrate the practical utility of the proposed algorithm in three Case studies with a broad set of network topologies, causal query types,  synthetic data generation approaches and causal query estimators. 
In particular, Case study 3 was based on observational experimental measurements of {\it E. Coli}, and was validated using interventional experimental measurements.
The causal generative approach was implemented with RStan~\citep{rstanInterface}.
More details, including model formulas, are in Appendix.
\subsection{Case study 1: The IGF signalling pathway}

\textbf{Input} \figref{fig:IGFgenerateSetOfRankedSubADMGs} (a) shows the full weighted ADMG of the insulin growth factor (IGF) signaling system, regulating growth and energy metabolism of a cell.
Variables are kinase proteins, and pointed/flat-headed edges represent the effect (increase/decrease) of the upstream kinase on the downstream kinase's activity. 
Assume that the costs of measuring each variable is 100 for $Ras$, 60 for $Raf$, 20 for $Mek$, 30 for $SOS$, and $\mathbb{W}=50$.
The causal query of interest is $E[Erk | do(Akt = 80)]$, and $N=100$.


\begin{figure*}[th]
\centering
\setlength{\tabcolsep}{1pt} 
\begin{tabular}{ccc}
\resizebox{7.2 cm}{!}{%
\tikz[]{
     \node[fill=yellow,draw, ellipse,align=left,label={[align=center, font=\Large]above:10}] (SARS) {\Large $\mathbf{SARS-}$\\\Large $\mathbf{Cov2}$};%
      \node[fill=yellow,draw,ellipse, right=of SARS,label={[align=center, font=\Large]above:1}] (ACE2) {\Large $\mathbf{ACE2}$};%
      \node[fill=yellow,draw,ellipse,right=of ACE2,label={[align=center, font=\Large]above:10}] (Ang) {\Large $\mathbf{AngII}$};%
      \node[fill=yellow,draw,ellipse,right=of Ang,label={[align=center, font=\Large]above:10}] (AGTR1) {\Large $\mathbf{AGTR1}$};%
      \node[fill=olive,draw,ellipse,right=of AGTR1,label={[align=center, font=\Large]above:1}] (ADAM17) {\Large $\mathbf{ADAM17}$};%
      \node[fill=yellow,draw,ellipse,below=of SARS,label={[align=center, font=\Large]above left:1}] (PRR) {\Large $\mathbf{PRR}$};%
      \node[fill=yellow,draw,ellipse,below=of ACE2,label={[align=center, font=\Large]above:1}] (Gefi) {\Large $\mathbf{Gefi}$};%
     \node[fill=yellow,draw,ellipse,below=of Ang,label={[align=center, font=\Large]above:1}] (EGF) {\Large $\mathbf{EGF}$};%
      \node[fill=olive,draw,ellipse,below=of ADAM17,label={[align=center, font=\Large]above left:1}] (sil) {\Large $\mathbf{sIL6R\alpha}$};%
      \node[fill=olive,draw,ellipse,below=of AGTR1,label={[align=center, font=\Large]above:1}] (tnf) {\Large $\mathbf{TNF\alpha}$};%
      \node[fill=green,draw,ellipse,below=of PRR,label={[align=center, font=\Large]above left:10}] (nfkb) {\Large $\mathbf{NFkB}$};%
      \node[fill=olive,ellipse,draw=purple, very thick, below=of EGF,label={[align=center, font=\Large]above right:1}] (EGFR) {\Large $\mathbf{EGFR}$};%
      \node[fill=olive,draw,ellipse,below=of tnf,align=left,label={[align=center, font=\Large]above:1}] (il6) {\Large $\mathbf{IL6-}$\\$\Large \mathbf{STAT3}$};%
      \node[fill=yellow,draw,ellipse,below=of sil,label={[align=center, font=\Large]above left:1}] (toci) {\Large $\mathbf{Toci}$};%
      \node[fill=green,draw,ellipse,below=of EGFR,label={[align=center, font=\Large]above:1}] (IL6) {\Large $\mathbf{IL6-AMP}$};%
      \node[fill=olive,ellipse,draw=orange, very thick,below=of IL6,label={[align=center, font=\Large]above right:1}, label={[align=center, font=\Huge] right:total weight : 52}] (Cytoke) {\Large $\mathbf{Cytoke}$};%
      \edge {SARS} {ACE2};
      \edge {ACE2} {Ang};
      \edge {Ang} {AGTR1};
      \edge {AGTR1} {ADAM17};
      \edge {SARS} {PRR};
      \edge {ADAM17} {EGF};
      \edge {ADAM17} {sil};
      \edge {ADAM17} {tnf};
      \edge {PRR} {nfkb};
      \edge {EGF}{EGFR};
      \edge {Gefi}{EGFR};
      \edge {sil}{il6};
      \edge {toci}{sil};
      \edge {tnf}{nfkb};
      \edge {EGFR}{nfkb};
      \edge {nfkb}{IL6};
      \edge {il6}{IL6};
      \edge {IL6}{Cytoke};
     
    \path[bidirected] (SARS) edge[bend left=30] (Ang);
    \path[bidirected] (PRR) edge[bend left=60] (nfkb);
    \path[bidirected] (EGF) edge[bend right=60] (EGFR);
    \path[bidirected] (tnf) edge[bend left=20] (EGFR);
    \path[bidirected] (il6) edge[bend left=20] (EGFR);
    \path[bidirected] (ADAM17) edge[bend left=60] (sil);
     } 
     } 
&
\resizebox{5 cm}{!}{%
\tikz[]{

      \node[fill=olive,draw,ellipse,label={[align=center, font=\Large]above:1}] (ADAM17) {\Large $\mathbf{ADAM17}$};%
      \node[fill=olive,draw,ellipse,below=of ADAM17,label={[align=center, font=\Large]above left:1}] (sil) {\Large $\mathbf{sIL6R\alpha}$};%
      \node[fill=olive,draw,ellipse,below=of sil,align=left,label={[align=center, font=\Large]above left:1}] (il6) {\Large $\mathbf{IL6-}$\\\Large $\mathbf{STAT3}$};%
      \node[fill=olive,draw,ellipse, left=of sil,label={[align=center, font=\Large]above:1}] (tnf) {\Large $\mathbf{TNF\alpha}$};%
      \node[fill=olive,ellipse,draw=purple, very thick, left=of tnf,label={[align=center, font=\Large]above:1}] (EGFR) {\Large $\mathbf{EGFR}$};%
      \node[fill=olive,ellipse,draw=orange, very thick,below left=of il6,label={[align=center, font=\Large] right:1}, label={[align=center, font=\huge, label distance=5mm] below:total weight : 6}] (Cytoke) {\Large $\mathbf{Cytoke}$};%
      \edge {ADAM17} {sil};
      \edge {ADAM17} {tnf};
      \edge {ADAM17} {EGFR};
      \edge {sil} {il6};
      \edge {tnf} {Cytoke};
      \edge {il6} {Cytoke};
      \edge {EGFR} {Cytoke};
    \path[bidirected] (tnf) edge[bend left=10] (EGFR);
    \path[bidirected] (il6) edge[bend left=10] (EGFR);
    \path[bidirected] (ADAM17) edge[bend left=50] (sil);
    \path[bidirected] (ADAM17) edge[bend left=98] (Cytoke);
}
} &
\resizebox{5 cm}{!}{%
\tikz[]{

      \node[fill=olive,draw,ellipse,label={[align=center, font=\Large]above:1}] (ADAM17) {\Large $\mathbf{ADAM17}$};%
      \node[fill=olive,draw,ellipse,below=of ADAM17,label={[align=center, font=\Large]above left:1}] (sil) {\Large $\mathbf{sIL6R\alpha}$};%
      \node[fill=olive,draw,ellipse,below=of sil,align=left,label={[align=center, font=\Large]above left:1}] (il6) {\Large $\mathbf{IL6-}$\\$\mathbf{STAT3}$};%
      \node[fill=olive,draw,ellipse, left=of sil,label={[align=center, font=\Large]above:1}] (tnf) {\Large $\mathbf{TNF\alpha}$};%
      \node[fill=olive,ellipse,draw=purple, very thick, left=of tnf,label={[align=center, font=\Large]above left:1}] (EGFR) {\Large $\mathbf{EGFR}$};%
      \node[fill=yellow,ellipse,draw, above right=of EGFR,label={[align=center, font=\Large]above:1}] (EGF) {\Large $\mathbf{EGF}$};%
      \node[fill=yellow,ellipse,draw, above=of EGFR,label={[align=center, font=\Large]above:1}] (Gefi) {\Large $\mathbf{Gefi}$};%
      \node[fill=olive,ellipse,draw=orange, very thick,below left=of il6,label={[align=center, font=\Large]right:1}, label={[align=center, font=\huge, label distance=5mm] below:total weight : 8}] (Cytoke) {\Large $\mathbf{Cytoke}$};%
      \edge {ADAM17} {sil};
      \edge {ADAM17} {tnf};
      \edge {ADAM17} {EGFR};
      \edge {EGF} {EGFR};
      \edge {sil} {il6};
      \edge {tnf} {Cytoke};
      \edge {il6} {Cytoke};
      \edge {EGFR} {Cytoke};
      \edge {Gefi} {EGFR};
    \path[bidirected] (tnf) edge[bend left=10] (EGFR);
    \path[bidirected] (EGF) edge[bend right=30] (EGFR);
    \path[bidirected] (il6) edge[bend left=10] (EGFR);
    \path[bidirected] (ADAM17) edge[bend left=50] (sil);
    \path[bidirected] (ADAM17) edge[bend left=98] (Cytoke);
}
}
     \\
(a) & (b) & (c)
\end{tabular}
\begin{tabular}{ccc}
\resizebox{5.2 cm}{!}{%
\tikz[]{
       \node[fill=yellow,draw,ellipse,label={[align=center, font=\Large]above:1}] (ACE2) {\Large $\mathbf{ACE2}$};%
       \node[fill=olive,draw,ellipse,right=of ACE2,label={[align=center, font=\Large]above:1}] (ADAM17) {\Large $\mathbf{ADAM17}$};
       \node[fill=yellow,draw,ellipse,below=of ACE2,label={[align=center, font=\Large]above:1}] (Gefi) {\Large $\mathbf{Gefi}$};%
       \node[fill=yellow,draw,ellipse,left=of Gefi,label={[align=center, font=\Large]above left:1}] (PRR) {\Large $\mathbf{PRR}$};%
       \node[fill=olive,draw,ellipse, right=of Gefi,label={[align=center, font=\Large]above right:1}] (tnf) {\Large $\mathbf{TNF\alpha}$};%
       \node[fill=olive,draw,ellipse,right=of tnf,label={[align=center, font=\Large]above right:1}] (sil) {\Large $\mathbf{sIL6R\alpha}$};%
       \node[fill=olive,ellipse,draw=purple, very thick, below =of Gefi,label={[align=center, font=\Large] right:1}] (EGFR) {\Large $\mathbf{EGFR}$};%
        \node[fill=green,draw,ellipse,below=of PRR,label={[align=center, font=\Large]above left:10}] (nfkb) {\Large $\mathbf{NFkB}$};%
      \node[fill=olive,draw,ellipse,right=of EGFR,align=left,label={[align=center, font=\Large]above:1}] (il6) {\Large $\mathbf{IL6-}$\\\Large $\mathbf{STAT3}$};%
       \node[fill=green,draw,ellipse,below=of EGFR,label={[align=center, font=\Large]above:1}] (IL6) {\Large $\mathbf{IL6-AMP}$};%
       \node[fill=olive,ellipse,draw=orange, very thick,below=of IL6,label={[align=center, font=\Large]above right:1}, label={[align=center, font=\Huge] right:total weight : 20}] (Cytoke) {\Large $\mathbf{Cytoke}$};%
       \edge {ACE2} {ADAM17};
       \path[] (ADAM17) edge[bend right=10] (EGFR);
       \edge {ADAM17} {tnf};
       \edge {ADAM17} {sil};
      \edge {PRR} {nfkb};
       \edge {Gefi}{EGFR};
       \edge {sil}{il6};
      \edge {tnf}{nfkb};
      \edge {EGFR}{nfkb};
      \edge {nfkb}{IL6};
      \edge {il6}{IL6};
       \edge {IL6}{Cytoke};
     
     \path[bidirected] (PRR) edge[bend left=60] (ADAM17);
      \path[bidirected] (ACE2) edge[bend left=20] (ADAM17);
     \path[bidirected] (PRR) edge[bend left=30] (ACE2);
     \path[bidirected] (tnf) edge[bend left=20] (EGFR);
     \path[bidirected] (il6) edge[bend left=20] (EGFR);
     \path[bidirected] (ADAM17) edge[bend left=60] (sil);
      } 
      } &
\raisebox{-0.04\height} {
\includegraphics[scale=0.17]{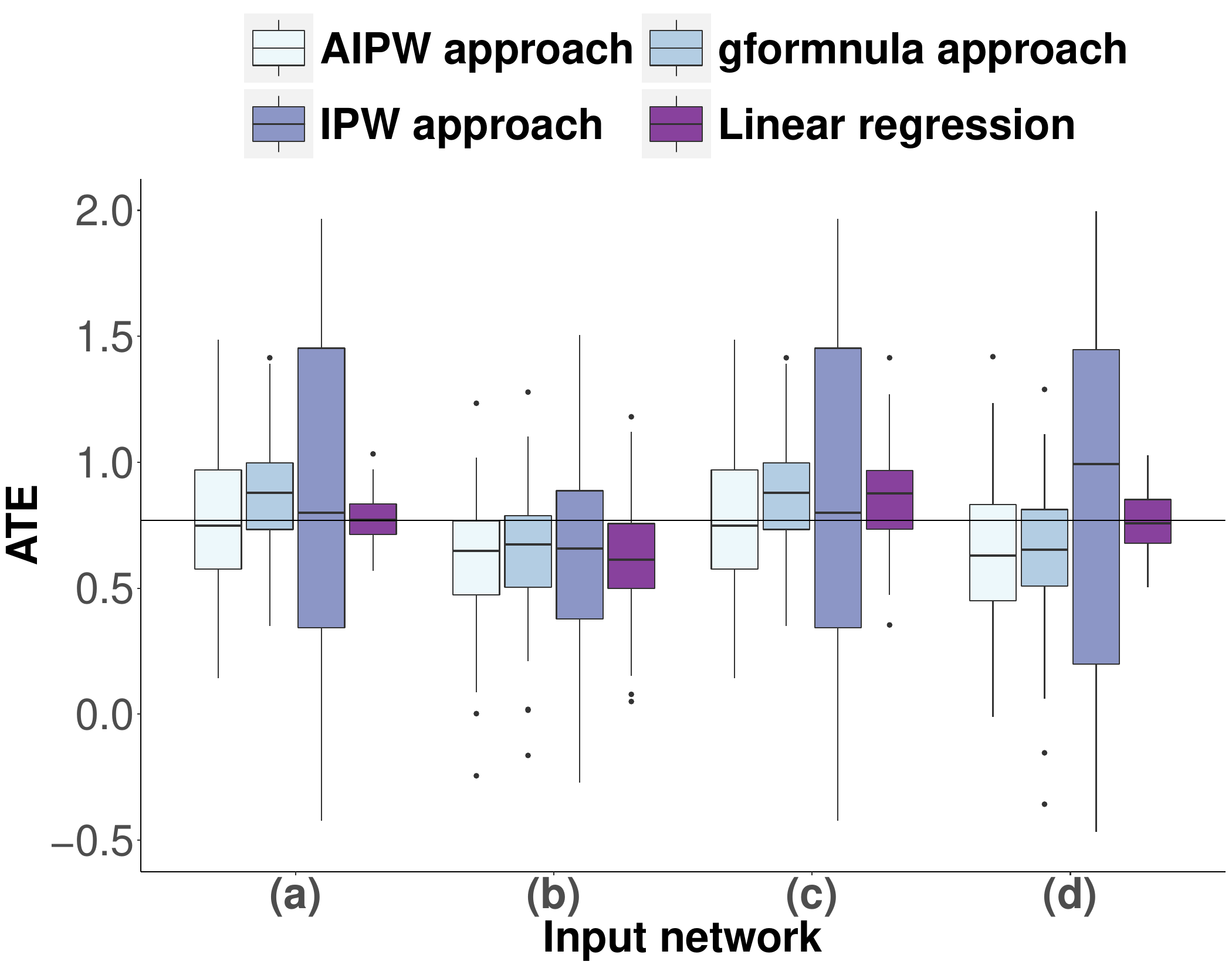} } &
\raisebox{0.97\height} {
\begin{small}
 \begin{tabular}{|c || c | c | c | c |} 
 \hline
  {} & \multicolumn{4}{|c|}{\textbf{{} Input networks } {} } \\[1mm]
 \cline{2-5}
 \begin{tabular}{@{}c@{}}
 \ \textbf{Nodes} \\
 \ \textbf{used by}
 \end{tabular} & 
 
 \textbf{(a)} &
 \textbf{ (b) } &
 \textbf{(c)} &
 \textbf{(d)}
 \\[1mm]
 \hline\hline
 
 \begin{tabular}{@{}c@{}}
 \ {} \\
 \ \textbf{ Linear } \\
 \ \textbf{ regression } \\
 \ {}
 \end{tabular}
 
 &
 
\multicolumn{4}{c|}{All input nodes}

\\ [1ex] 
  
 \hline
 \begin{tabular}{@{}c@{}}
 \ {} \\
 \ \textbf{AIPW} \\
 \ \textbf{IPW} \\
 \ \textbf{gformula} \\
 \ {}
 \end{tabular}

&
 
 \begin{tabular}{@{}c@{}}
 \ {} \\
 \ {olive } \\
 \ {Gefi } \\
 \ {EGF }
 \end{tabular}

&
 
\multicolumn{2}{c|}{
 \begin{tabular}{@{}c@{}}
 \ {} \\
 \ {All input } \\
 \ {nodes}
 \end{tabular}
} 

&

 \begin{tabular}{@{}c@{}}
 \ {} \\
 \ \text{olive } \\
 \ Gefi
 \end{tabular}

\\
\hline
\end{tabular} 
\end{small}
\label{tables:Covid}
}
\\
(d) & (e) & (f)
\end{tabular}
    \caption{\small 
    {\bf Case study 2.}
    \textbf{(a)} Full weighted ADMG. Node colors are as in \figref{fig:IGFgenerateSetOfRankedSubADMGs}.
    $EGFR$ is the treatment and $Cytokine Storm$ is the effect. Bi-directed edges indicate presence of latent variables.
    (b)-(d) Example Sub-ADMGs.
    Simplification rules in \secref{sec:GraphNotCausalInf} lead \algref{alg:generateSetOfRankedSubADMGs} to convert the latent variables into exogenous variables, shown with bi-directed edges.
    (e) Sampling distribution of $ATE=E[CS|do(EGFR=1)] - E[CS|do(EGFR=0)]$ over $K=100$ simulated data, $N=1000$. 
    The horizontal line is the true ATE from interventional data.
    The apparent bias was due to the approximations in the model assumptions, and to the asymptotic nature of the estimators. 
    (f) Variables used by different query estimators. 
    \label{fig:CovidCaseStudy}
    }
\end{figure*}

\textbf{Step 1: Simulating future data sets}  Since IGF dynamics is well characterized in form of stochastic differential equations (SDE), we did not use any historical data. Instead, we generated $K=20$ observational
datasets by simulating from the SDE. 
We set the initial amount of each protein to 100, and generated subsequent observations via the Gillespie algorithm~\cite{gillespie1977exact} with the \textit{smfsb}~\cite{wilkinson2018package} R package. 
To evaluate the true value of the query, we simulated interventional data while fixing $Akt=80$.

\textbf{Step 2: Exploring sub-ADMGs} \figref{fig:IGFgenerateSetOfRankedSubADMGs}(a)-(d) shows some sub-ADMGs explored by~\algref{alg:generateSetOfRankedSubADMGs}. Since $\{PI3K\}$ was a valid adjustment set, it always remained in the search space together with the treatment ($Akt$) and the effect ($Erk$). $Ras$ and $Raf$ were absent as their cost exceeded $\mathbb{W}$.

\textbf{Step 3: Estimating causal query} The non-parametric and semi-parametric estimation approaches were not applicable to the continuous treatment $Akt$. Hence, we used the causal generative and the linear regression estimation approaches. 

\textbf{Output and conclusions}
\figref{fig:IGFgenerateSetOfRankedSubADMGs}(e) summarizes the queries estimated over $K=20$ simulated observational datasets. 
We can make several conclusions.
1) Including a single mediator in \figref{fig:IGFgenerateSetOfRankedSubADMGs}(c) to the adjustment set of olive nodes in \figref{fig:IGFgenerateSetOfRankedSubADMGs}(b) reduced the variance of the causal query estimator by over 50\% (\tableref{tables:IGF});
2) Using the full network in \figref{fig:IGFgenerateSetOfRankedSubADMGs}(a) was suboptimal.
Other sub-ADMGs had a comparable variance and lower cost (\tableref{tables:IGF});
3) The choice of estimator and of $N$ affected the ranking of sub-ADMGs (\tableref{tables:IGF}).



\begin{figure*}[t]
\begin{center}
\setlength{\tabcolsep}{1pt} 

\begin{tabular}{l}
\includegraphics[scale=0.44]{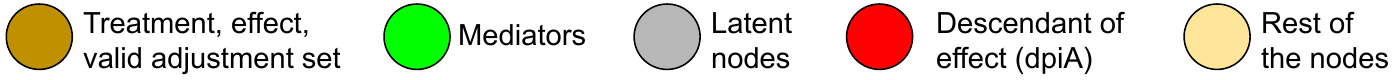} 
\end{tabular}

\begin{tabular}{ccc}
\resizebox{7.5 cm}{!}{%
\tikz[]{
     \node (rpoD) at (5,0)   [fill=olive,draw,ellipse,minimum size=1.6cm,label={[align=center, font=\Large]below right:1}]  {\huge $\mathbf{rpoD}$};%
     \node (chiX) at (-8,-1) [obs,ellipse,minimum size=1.6cm,label={[align=center, font=\Large]above right:1}]  {\huge $\mathbf{chiX}$};%
     \node (crp)  at (-4,-1) [fill=olive,draw,ellipse,minimum size=1.6cm,label={[align=center, font=\Large]above right:1}]  {\huge $\mathbf{crp}$};%
     \node (oxyR) at (-0.5,-3) [fill=yellow,draw,ellipse,minimum size=1.6cm,label={[align=center, font=\Large]above:1}]  {\huge $\mathbf{oxyR}$};%
     \node (fur)  at (-8,-4) [fill=olive,draw=purple,very thick,ellipse,minimum size=1.6cm,label={[align=center, font=\Large]above:1}] {\huge $\mathbf{fur}$};%
     \node (fnr)  at (0,-5) [fill=green,draw,ellipse,minimum size=1.6cm,label={[align=center, font=\Large]right:1}] {\huge $\mathbf{fnr}$};%
     \node (narL)  at (9,-7) [fill=green,draw,ellipse,minimum size=1.6cm,label={[align=center, font=\Large]above:1}] {\huge $\mathbf{narL}$};%
     \node (dcuR)  at (1,-8) [fill=green,draw,ellipse,minimum size=1.6cm,label={[align=center, font=\Large]above right:1}] {\huge $\mathbf{dcuR}$};%
     \node (dpiA)  at (-3,-10) [fill=olive,draw=orange,very thick,ellipse,minimum size=1.6cm,label={[align=center, font=\Large]above right:1}] {\huge $\mathbf{dpiA}$};%
     \node (appY)  at (6.5,-11) [fill=red,draw,ellipse,minimum size=1.6cm,label={[align=center, font=\Large]above right:1}] {\huge $\mathbf{appY}$};%
     \node (aspC)  at (-10,-12.3) [fill=red,draw,ellipse,minimum size=1.6cm,label={[align=center, font=\Large]above:1}] {\huge $\mathbf{aspC}$};%
     \node (citX)  at (-6.5,-12.3) [obs,ellipse,minimum size=1.6cm,label={[align=center, font=\Large]above left:1}] {\huge $\mathbf{citX}$};%
     \node (dpiB)  at (.1,-12.8) [fill=red,draw,ellipse,minimum size=1.6cm,label={[align=center, font=\Large]right:1}, label={[align=center, font=\Huge, label distance=5mm] below:total weight : 14}] {\huge $\mathbf{dpiB}$};%
     \node (hyaA)  at (4,-12.5) [fill=red,draw,ellipse,minimum size=1.6cm,label={[align=center, font=\Large]above left:1}] {\huge $\mathbf{hyaA}$};%
     \node (hyaB)  at (9,-12.5) [fill=red,draw,ellipse,minimum size=1.6cm,label={[align=center, font=\Large]above right:1}] {\huge $\mathbf{hyaB}$};%
     \node (hyaF)  at (13,-12.5) [fill=red,draw,ellipse,minimum size=1.6cm,label={[align=center, font=\Large]above right:1}] {\huge $\mathbf{hyaF}$};%
       \path[] (rpoD) edge[bend right=10] (chiX);
       \path[] (rpoD) edge[bend left=10] (dpiB);
       \path[] (narL) edge[bend right=17] (hyaA);
       \path[] (fnr) edge[bend right=12] (dpiB);
       \path[] (fnr) edge[bend right=11] (citX);
       \edge {rpoD} {crp};
       \edge {crp} {fur};
       \edge {rpoD} {fur};
       \edge {oxyR} {fur};
       \edge {rpoD} {oxyR};
       \edge {crp} {oxyR};
       \edge {fur} {fnr};
       \edge {rpoD} {fnr};
       \edge {fnr}{narL};
       \edge {rpoD}{narL};
       \edge {fnr}{dcuR};
       \edge {crp}{dcuR};
       \edge {rpoD}{dcuR};
       \edge {narL}{dcuR};
       \edge {chiX}{dpiA};
       \edge {crp}{dpiA};
       \edge {fnr}{dpiA};
       \edge {dcuR}{dpiA};
       \edge {narL}{dpiA};
       \edge {dpiA}{appY};
       \edge {rpoD}{appY};
       \edge {fur}{aspC};
       \edge {fnr}{aspC};
       \edge {crp}{citX};
       \edge {dpiA}{citX};
       \edge {narL}{citX};
       \edge {dpiA}{dpiB};
       \edge {crp}{dpiB};
       \edge {dcuR}{dpiB};
       \edge {narL}{dpiB};
       \edge {citX}{dpiB};
       \edge {rpoD}{hyaA};
       \edge {appY}{hyaA};
       \edge {appY}{hyaB};
       \edge {rpoD}{hyaB};
       \edge {narL}{hyaB};
       \edge {appY}{hyaF};
       \edge {narL}{hyaF};
       \path[] (rpoD) edge[bend left=30] (hyaF);
     

     } 
} &
\resizebox{4.7 cm}{!}{%
\tikz[]{
     \node (rpoD) at (5,0)   [fill=olive,draw,ellipse,minimum size=1.6cm,label={[align=center, font=\Large]above:1}]  {\huge $\mathbf{rpoD}$};%
     \node (crp)  at (-4,-1) [fill=olive,draw,ellipse,minimum size=1.6cm,label={[align=center, font=\Large]left:1}]  {\huge $\mathbf{crp}$};%
     \node (oxyR) at (-0.5,-3) [fill=yellow,draw,ellipse,minimum size=1.6cm,label={[align=center, font=\Large]above:1}]  {\huge $\mathbf{oxyR}$};%
     \node (fur)  at (-8,-4) [fill=olive,draw=purple,very thick,ellipse,minimum size=1.6cm,label={[align=center, font=\Large]above:1}] {\huge $\mathbf{fur}$};%
     \node (fnr)  at (0,-5) [fill=green,draw,ellipse,minimum size=1.6cm,label={[align=center, font=\Large]above:1}] {\huge $\mathbf{fnr}$};%
     \node (narL)  at (5,-7) [fill=green,draw,ellipse,minimum size=1.6cm,label={[align=center, font=\Large]above right:1}] {\huge $\mathbf{narL}$};%
     \node (dcuR)  at (0.5,-8) [fill=green,draw,ellipse,minimum size=1.6cm,label={[align=center, font=\Large]above right:1}] {\huge $\mathbf{dcuR}$};%
     \node (dpiA)  at (-1,-10) [fill=olive,draw=orange,very thick,ellipse,minimum size=1.6cm,label={[align=center, font=\Large]left:1}, label={[align=center, font=\Huge, label distance=5mm] below:total weight : 8}] {\huge $\mathbf{dpiA}$};%

      \edge {rpoD} {crp};
      \edge {crp} {fur};
      \edge {rpoD} {fur};
      \edge {oxyR} {fur};
      \edge {rpoD} {oxyR};
      \edge {crp} {oxyR};
      \edge {fur} {fnr};
      \edge {rpoD} {fnr};
      \edge {fnr}{narL};
      \edge {rpoD}{narL};
      \edge {fnr}{dcuR};
      \edge {crp}{dcuR};
      \edge {rpoD}{dcuR};
      \edge {narL}{dcuR};
      \edge {crp}{dpiA};
      \path[] (fnr) edge[bend right=20] (dpiA);
      \edge {dcuR}{dpiA};
      \edge {narL}{dpiA};
     

     } 
} &
\resizebox{2.7 cm}{!}{%
\tikz[]{
     \node (rpoD) at (-1,-1)   [fill=olive,draw,ellipse,minimum size=1.6cm,label={[align=center, font=\Large]above:1}]  {\huge $\mathbf{rpoD}$};%
     \node (crp)  at (-6,-2) [fill=olive,draw,ellipse,minimum size=1.6cm,label={[align=center, font=\Large]left:1}]  {\huge $\mathbf{crp}$};
     \node (fur)  at (-8,-5) [fill=olive,draw=purple,very thick,ellipse,minimum size=1.6cm,label={[align=center, font=\Large]above:1}] {\huge $\mathbf{fur}$};%
     \node (dpiA)  at (-6,-10) [fill=olive,draw=orange,very thick,ellipse,minimum size=1.6cm,label={[align=center, font=\Large]right:1}, label={[align=center, font=\Huge, label distance=5mm] below:total weight : 4}] {\huge $\mathbf{dpiA}$};%

      \edge {rpoD} {crp};
      \edge {rpoD} {dpiA};
      \edge {crp} {fur};
      \edge {rpoD} {fur};
      \edge {crp}{dpiA};
      \edge {fur}{dpiA};
     

     } 
} \\
(a) & (b) & (c)
\end{tabular}

\begin{tabular}{ccc}
\resizebox{2.6 cm}{!}{%
\tikz[]{
     \node (rpoD) at (0,-1)   [fill=olive,draw,ellipse,minimum size=1.6cm,label={[align=center, font=\Large]above:1}]  {\huge $\mathbf{rpoD}$};%
     \node (crp)  at (-4,-2) [fill=olive,draw,ellipse,minimum size=1.6cm,label={[align=center, font=\Large]above:1}]  {\huge $\mathbf{crp}$};%
     \node (fur)  at (-6,-5) [fill=olive,draw=purple,very thick,ellipse,minimum size=1.6cm,label={[align=center, font=\Large]above:1}] {\huge $\mathbf{fur}$};%
     \node (fnr)  at (0.5,-8) [fill=green,draw,ellipse,minimum size=1.6cm,label={[align=center, font=\Large]above right:1}] {\huge $\mathbf{fnr}$};%
     \node (dpiA)  at (-3,-10) [fill=olive,draw=orange,very thick,ellipse,minimum size=1.6cm,label={[align=center, font=\Large]left:1}, label={[align=center, font=\Huge, label distance=8mm] below:total weight : 5}] {\huge $\mathbf{dpiA}$};%

      \edge {rpoD} {crp};
      \edge {crp} {fur};
      \edge {rpoD} {fur};
      \edge {rpoD} {dpiA};
      \edge {fur}{fnr};
      \edge {crp}{dcuR};
      \edge {rpoD}{dcuR};
      \edge {crp}{dpiA};
      \edge {fnr}{dpiA};
     } 
} &

\resizebox{4.3 cm}{!}{%
\tikz[]{
     \node (rpoD) at (5,0)   [fill=olive,draw,ellipse,minimum size=1.6cm,label={[align=center, font=\Large]above:1}]  {\huge $\mathbf{rpoD}$};%
     \node (crp)  at (-4,-1) [fill=olive,draw,ellipse,minimum size=1.6cm,label={[align=center, font=\Large]above:1}]  {\huge $\mathbf{crp}$};%
     \node (fur)  at (-6,-4) [fill=olive,draw=purple,very thick,ellipse,minimum size=1.6cm,label={[align=center, font=\Large]above:1}] {\huge $\mathbf{fur}$};%
     \node (fnr)  at (0,-5) [fill=green,draw,ellipse,minimum size=1.6cm,label={[align=center, font=\Large]above:1}] {\huge $\mathbf{fnr}$};%
     \node (narL)  at (6,-7) [fill=green,draw,ellipse,minimum size=1.6cm,label={[align=center, font=\Large]above left:1}] {\huge $\mathbf{narL}$};%
     \node (dcuR)  at (0.5,-8) [fill=green,draw,ellipse,minimum size=1.6cm,label={[align=center, font=\Large]above right:1}] {\huge $\mathbf{dcuR}$};%
     \node (dpiA)  at (-1,-10) [fill=olive,draw=orange,very thick,ellipse,minimum size=1.6cm,label={[align=center, font=\Large]left:1}, label={[align=center, font=\Huge, label distance=4mm] below:total weight : 7}] {\huge $\mathbf{dpiA}$};%

      \edge {rpoD} {crp};
      \edge {crp} {fur};
      \edge {rpoD} {fur};
      \edge {fur} {fnr};
      \edge {rpoD} {fnr};
      \edge {fnr}{narL};
      \edge {rpoD}{narL};
      \edge {fnr}{dcuR};
      \edge {crp}{dcuR};
      \edge {rpoD}{dcuR};
      \edge {narL}{dcuR};
      \edge {crp}{dpiA};
       \path[] (fnr) edge[bend right=20] (dpiA);
      \edge {dcuR}{dpiA};
      \edge {narL}{dpiA};
     

     } 
} &
\raisebox{-0.02\height} {
\includegraphics[scale=0.17]{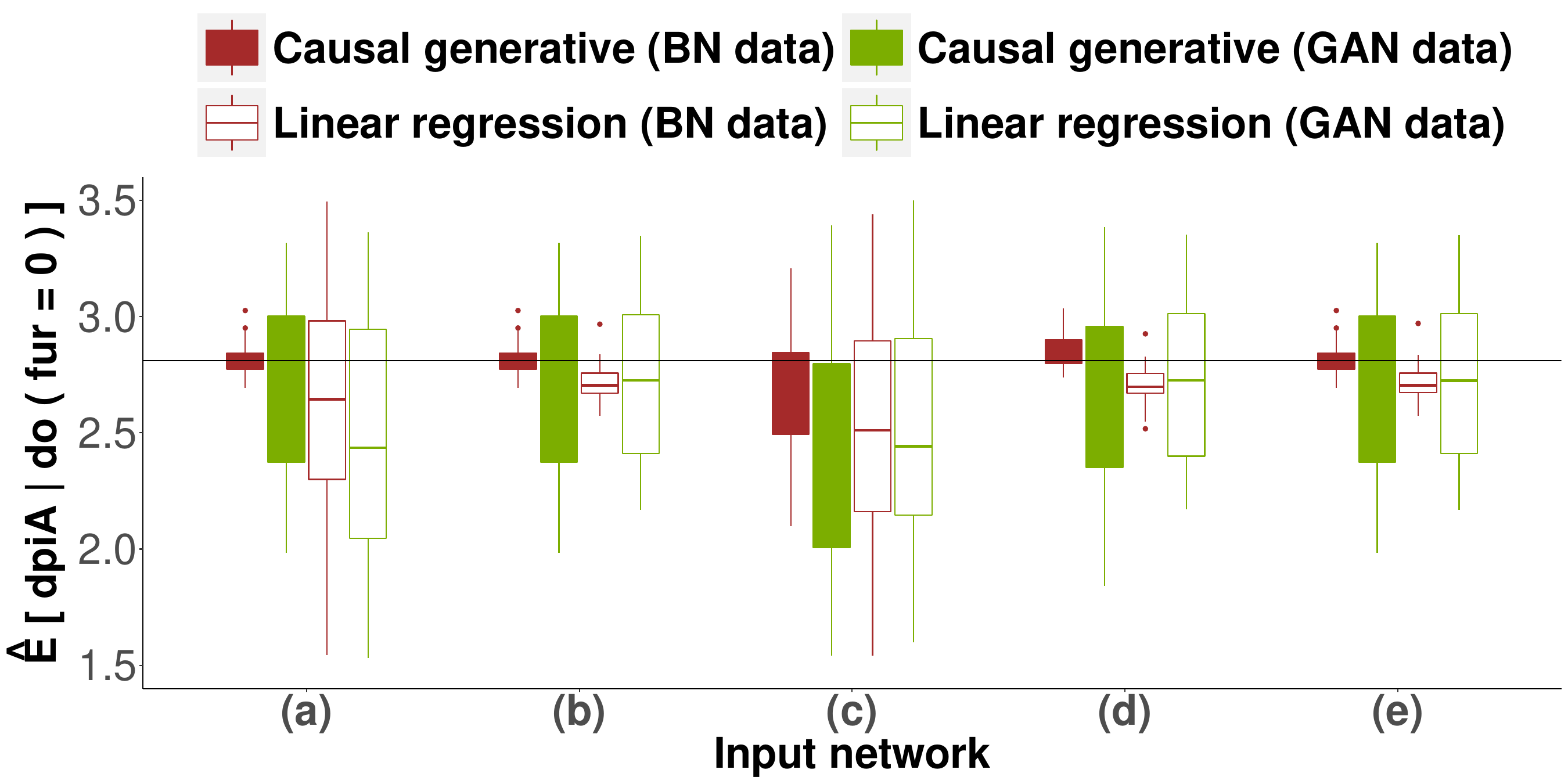} }
\\
(d) & (e) & (f)
\end{tabular}

\end{center}
    \caption{\small \textbf{Case study 3.} 
    (a) Full weighted ADMG. $fur$ is the treatment and $dpiA$ is the effect. 
    (b) \algref{alg:generateSetOfRankedSubADMGs} removed the red and grey nodes from the search space.
    (c)-(e) Example sub-ADMGs. 
    Simplification rules in \secref{sec:GraphNotCausalInf} lead \algref{alg:generateSetOfRankedSubADMGs} removing the latent variables.
    (f) Sampling distribution of $Q_{fur}=P(dpiA | do(fur=0))$ estimation over $K=100$ simulated data, $N=1000$. 
    The horizontal line is true value obtained experimentally, upon an intervention.
    The apparent bias was due to the approximations in the model assumptions, and to the asymptotic nature of the estimators. 
    \label{fig:experimentalCaseStudy}}
\end{figure*}
\subsection{Case study 2: The SARS-CoV-2 model}
\label{section:SyntCaseCovidModel}
\textbf{Input} The full weighted ADMG in \figref{fig:CovidCaseStudy}(a)
models activation of Cytokine Release Syndrome (cytokine storm), known to cause tissue damage in severely ill SARS-CoV-2 patients \cite{ulhaq_2020}. The simultaneous activation of the nuclear factor kappa-light-chain-enhancer of activated B cells (NF-$\kappa$B) and Interleukin 6-STAT3 Complex (IL6-STAT3) initiates a positive feedback loop known as Interleukin 6 Amplifier (IL6-AMP), which in turn activates Cytokine Storm \cite{hirano_2020}. 
Gefitinib (Gefi) iis an immunosuppressive drug that blocks epidermal growth factor receptor (EGFR),, and prevents the cytokine storm.  
The causal query examines the $ATE=E[CS|do(EGFR=1)] - E[CS|do(EGFR=0)]$. 
$\mathbb{W}=\infty$, and $N = 1000$.

\textbf{Step 1: Simulating future data sets} For this case study, the generation of synthetic data was motivated by common biological practice.
Simple biomolecular reactions were modeled with Hill function \cite{alon2019introduction} as $\mathcal{N}(\frac{100}{1 + exp(\mathbf{\theta}^{\prime}  Pa(X) + \theta_0)}, \sigma_X)$, where $Pa(X)$ is a $q \times 1$ vector of measurements on the parent of $X$, $\mathbf{\theta}'$ is a $1 \times q$ vector of  parameters, and $\theta_0$ is a scalar. 
Probability distributions of the exogenous variables were simulated as $\mathcal{N}(\mu_r, \sigma_r)$.
The treatment $EGFR$ was binary.
Interventional data were generated similarly, while fixing $EGFR$ to 1 and 0, and were used to define the true value of the query. $K=100$.

\textbf{Step 2: Exploring sub-ADMGs} \figref{fig:CovidCaseStudy}(a)-(d) shows some sub-ADMGs explored by~\algref{alg:generateSetOfRankedSubADMGs}.
The  adjustment set, the treatment and the effect (olive nodes) were always present.

\textbf{Step 3: Estimating causal query} Since the treatment is single and binary, we considered the semi- and non-parametric approaches in \texttt{Ananke} (AIPW, IPW, and gformula), and the linear regression. RStan was unable to fit a causal generative approach with a discrete variable.

\textbf{Output and conclusions} 
\figref{fig:CovidCaseStudy}(e) summarizes the queries estimated over $K=100$ simulated observational data sets. We can make several conclusions.  
1) For linear regression, the full network in \figref{fig:CovidCaseStudy}(a) and the sub-network in \figref{fig:CovidCaseStudy}(d) had similar variances, but at different cost.
2) For linear regression, compared to using the adjustment set alone in \figref{fig:CovidCaseStudy}(b), including mediators in \figref{fig:CovidCaseStudy}(d) reduced the variance of the query estimation. However, this was not the case for AIPW, IPW, and gformula estimators, as they were constrained to certain input variables regardless of the input (\figref{fig:CovidCaseStudy}(f)).
3) For AIPW, IPW, and gformula, the full network in \figref{fig:CovidCaseStudy}(a) and the sub-network in \figref{fig:CovidCaseStudy}(c) had exact same results but at different cost.




\subsection{Case study 3: The Escherichia coli K-12 transcriptional motif}

\textbf{Input} The full weighted ADMG in \figref{fig:experimentalCaseStudy} (a) is a transcriptional regulatory network motif of {\it E. Coli}. 
The network was obtained from the EcoCyc database \cite{keseler2021ecocyc}. All the weights were set to 1. The query of interest is $Q_{fur}=P(dpiA | do(fur=0))$. $\mathbb{W}=\infty$, and $N = 262$.

\textbf{Step 1: Simulating future data sets} We used 262 RNA-seq normalized expression profiles of {\it E. Coli} K-12 MG1655 and BW25113 across 154 unique experimental conditions from the PRECISE database \citep{sastry2019escherichia}. The dataset also included interventional data with $fur = 0$. It was used to evaluate the accuracy of the estimators. Based on these historical data, we generated two synthetic datasets. 
The first was generated from a Bayesian Network, trained on the experimental data using parametrized generalized linear models (GLMs). 
At each node the GLMs had a Normal or Gamma distribution with identity/log link functions. Bayesian Information Criterion (BIC) was used to guide feature engineering/transformations.
Once the model was trained, new samples were generated by forward simulation. 
The second synthetic dataset was generated from Tabular GAN \cite{xu2019modeling}. We prepossessed the data as described in~\cite{ashrapov2020tabular} with default hyperparameters. 

\noindent \textbf{Step 2: Exploring sub-ADMGs} The adjustment set are the olive nodes in \figref{fig:experimentalCaseStudy} excluding the treatment and effect.
Selected sub-ADMGs from the whole set of sub-ADMGs are shown in \figref{fig:experimentalCaseStudy} (c)-(e) with their corresponding measurement costs. 
\figref{fig:experimentalCaseStudy} (b) shows the simplified ADMG of the full network in \figref{fig:experimentalCaseStudy} (a) where the gray, and red nodes are removed. 
Sub-ADMG in \figref{fig:experimentalCaseStudy} (c) only consist of the treatment, effect, and the adjustment set.
Sub-ADMG in \figref{fig:experimentalCaseStudy} (d) contains a single mediator.
Sub-ADMG in \figref{fig:experimentalCaseStudy} (e) contains all the mediators.

\noindent \textbf{Step 3: Estimating causal query} Since the treatment $fur$ was continuous, we considered the causal generative and linear regression estimators. 

\noindent \textbf{Output and conclusions} \figref{fig:experimentalCaseStudy} (f) summarizes the queries estimated over $K=100$ simulated observational data sets. We can make several conclusions.
1) Since causal generative model did not use all the input nodes, using the full network in \figref{fig:experimentalCaseStudy}(a) had the exact same result as in \figref{fig:experimentalCaseStudy}(b) and (e). However, the full network had a larger cost.
2) For linear regression, using the full network in \figref{fig:experimentalCaseStudy}(a) increased the variance and slightly biased the results as compared to most other sub-ADMGs. 
3) Compared to only using the treatment, effect and adjustment set (olive nodes) in \figref{fig:experimentalCaseStudy}(c), including a single mediator (\figref{fig:experimentalCaseStudy} (d)) reduced the variance by over 50\% for most estimators;
4) Including all the mediators (\figref{fig:experimentalCaseStudy}(e) slightly decreased the variance compared to the subnetwork with a single mediator in \figref{fig:experimentalCaseStudy}(d), but at extra cost;
5) Different simulation strategies lead to same ranking of sub-ADMGs, but with different variances.

\subsection{Summary of the conclusions of the Case studies}

\noindent \textbf{Expanding the set of measured variables beyond the adjustment was beneficial in presence of mediators.}
Lemma 1 stated theoretically, and all the case studies showed empirically, that expanding the adjustment set with at least one mediator reduced the variance of the causal query estimator. The benefits of including multiple mediators were Case study-specific.

\noindent \textbf{Measuring a subset of the variables was often more effective than measuring the entire network.} In illustrative example and Case study 3, using all the variables biased the causal query estimator. In Case studies 1 and 2, well-chosen subsets of variables resulted in similar or smaller variances of the estimators, but at lower cost.

\noindent \textbf{Experimental design depended on the choice of the query estimation method.} The different restrictions on the input variables, and the different functional forms of the estimators impacted the variances associated with subsets of variables, as illustrated in Case studies 1 and 2.

\noindent \textbf{Experimental design depended on the sample size.} The sample size $N$ affected the relative performance of the subsets of variables and of the estimators in Case study 1.

\section{Discussion}


We proposed an approach for selecting optimal subsets of variables, to quantify in future observational studies with the purpose of causal query estimation. 
A limitation of the proposed approach is the requirement of a known biomolecular network. 
We overcame this limitation in part by working with ADMGs, which allow us to misspecify the latent variables as long as the observed variables are represented accurately. 
The correctness of the structure over observed variables can be checked with the falsification module in $Y_0$~\cite{y0}.
Another potential limitation is the accuracy of the generated synthetic data, which can be addressed by sensitivity analysis.
Finally, although the estimators supported by the proposed approach are (asymptotically) unbiased, they may be biased in practice. This is due to violations of model assumptions, and to the relatively small sample size. Simulating interventional data, as in the Case studies in this manuscript, helps diagnose and account for this issue in the choice of a sub-ADMG.

Future directions of this research include a closer integration with simulated interventional data. This will allow us to expand the scope of causal queries, check their idenifiability, e.g. with the G-ID algorithm~\cite{pmlr-v115-lee20b}, and increase their practical use.

\section*{Funding}
JZ is supported by the PNNL Directed R\&D-funded Data-Model Convergence Initiative. PNNL is operated for the DOE by Battelle Memorial Institute under Contract DE-AC05-76RLO1830.
CTH is supported by the DARPA Young Faculty Award W911NF2010255 (PI: Benjamin M. Gyori).
KS is supported by Muscular Dystrophy Association MDA Award \#574137.
OV acknowledges the support of NSF-BIO/DBI 1759736, NSF-BIO/DBI 1950412, NIH-NLM-R01 1R01LM013115 and of the Chan-Zuckerberg foundation.

\bibliographystyle{unsrtnat}
\bibliography{references}  

\clearpage
\appendix
{\Large \bf Appendix}
\section{Algorithms}
We provide details of the functions used in \algref{alg:generateSetOfRankedSubADMGs}.

\algref{alg:generateValidAdjustmentSet} takes as input an ADMG, treatment, effect, and the upper limit on the total weight and outputs a valid adjustment set with minimum weight such that its total cost does not exceed the upper limit total weight and it contains all the good neutral controls.

\algref{alg:generateValidSubsets} takes as input a set of weighted nodes and produces their powerset conditioned on having sum of weights in each subset to be less than a given threshold. The subsets are recursively constructed by calling  \algref{alg:generateValidSubsetsHelper}.

\algref{alg:generateSubADMG} takes as input an ADMG, and the set of variables that the experimenter decides to measure, and it outputs the corresponding sub-ADMG that contains only the new measure set by applying the simplification rules in \secref{sec:GraphNotCausalInf}.




\begin{algorithm}[th]
\small
\caption{generateValidAdjustmentSet}
\label{alg:generateValidAdjustmentSet}
\DontPrintSemicolon
\SetAlgoLined
\KwInput{$\mathcal{G} = (\mathbf{V}, \mathbf{E}_d, \mathbf{E}_b, \mathbf{W})$: Weighted ADMG\\
$~~~~~~~~~~~$ $T, Y$: Treatment and effect\\
$~~~~~~~~~~~$ $\mathbb{W}$: Upper limit on the total weight
}

\KwOutput{Valid adjustment set with minimum weight that includes the variables that possibly decrease the variance of the query estimator (good neutral controls)}

\vspace{0.2cm}
\hrule 
\vspace{0.2cm}

\tcp{All valid adjustment sets from the dagitty software}
$\mathbf{A} =$  adjustmentSets($\mathcal{G}$, $T$, $Y$)

minAdjSetCost = Inf

selectedAdjustmentSet = \{\}

forbiddenSet = $de(cp(T,Y,\mathcal{G})) \cup T$

\tcp{Good neutral control set}
gnControlSet = $Pa(cp(T,Y,\mathcal{G})) \backslash$forbiddenSet

\For{a $\in$ $\mathbf{A}$}{

    $\mathbb{W}_a = \textnormal{sumWeights(a)}$
    
    \If{$\mathbb{W}_a > \mathbb{W}$}{continue}

    \If{$\mathbb{W}_a <$ minAdjSetCost and $a \subseteq$ gnControlSet}{
    
        minAdjSetCost = $\mathbb{W}_a$
        
        selectedAdjustmentSet = a
        
    }

}

\textbf{return} selectedAdjustmentSet

\end{algorithm}


\begin{algorithm}[b]
\small
\caption{getValidNodeSubsets}
\label{alg:generateValidSubsets}
\DontPrintSemicolon
\SetAlgoLined
\KwInput{$\mathcal{G} = (\mathbf{V}, \mathbf{E}_d, \mathbf{E}_b, \mathbf{W})$: Weighted ADMG\\
$~~~~~~~~~~~$ $\mathbf{A}$: A valid adjustment set\\
$~~~~~~~~~~~$ $T, Y$: Treatment and effect\\
$~~~~~~~~~~~$ $\mathbb{W}$: Upper limit on the total weight
}

\KwOutput{Subsets of nodes in the given ADMG which has sum of weights less than or equal to $\mathbb{W}$ and also include $T$, $Y$, and the nodes in $A$.}

\vspace{0.2cm}
\hrule 
\vspace{0.2cm}


requiredNodes $\leftarrow \mathbf{A} \cup T \cup Y$ {\color{brown} \ttfamily \scriptsize //Olive nodes} \label{line:requiredCol}

optionalNodes $\leftarrow \mathbf{V} \backslash \{\text{requiredNodes} \}$ {\color{brown} \ttfamily \scriptsize //Yellow and green nodes} \label{line:optionalCol}

$\mathbf{result}$ = []

\tcp{Recursively add subsets to the result.}

$\mathbf{getValidNodeSubsetsHelper}$(requiredNodes, optionalNodes, $\mathbb{W}$, 0, [], result)

\textbf{return} $\mathbf{result}$

\end{algorithm}


\begin{algorithm}[t]
\small
\caption{getValidNodeSubsetsHelper}
\label{alg:generateValidSubsetsHelper}
\DontPrintSemicolon
\SetAlgoLined
\KwInput{requiredNodes : Set of nodes present in each subset \\
$~~~~~~~~~~~$ optionalNodes: Set of optional nodes in a subset \\
$~~~~~~~~~~~$ $\mathbb{W}$: Total weight limit \\
$~~~~~~~~~~~$ $\mathbf{idx}$: Current index \\
$~~~~~~~~~~~$ $\mathbf{subset}$: Current subset \\
$~~~~~~~~~~~$ $\mathbf{result}$: Power set
}

\KwOutput{Recursively adds subsets of optionalNodes to the result. requiredNodes are also added to each subset.}

\vspace{0.2cm}
\hrule 
\vspace{0.2cm}

\If{sum(subset.weight) $+$ sum(requiredNodes.weight)  $>$ $\mathbb{W}$}{
\textbf{return}
}

\If{idx $==$ optionalNodes.length}{
subset.add(requiredNodes)

$\mathbf{result}$.add(subset))
}

$\mathbf{getValidNodeSubsetsHelper}$(requiredNodes, optionalNodes, $\mathbb{W}$, idx+1, subset.copy(), result)

subset.add(optionalNodes[idx])

$\mathbf{generateValidSubsetsHelper}$(requiredNodes, optionalNodes, $\mathbb{W}$, idx+1, subset.copy(), result)

\end{algorithm}



\begin{algorithm}[t]
\small
\caption{generateSubADMG}
\label{alg:generateSubADMG}
\DontPrintSemicolon
\SetAlgoLined
\KwInput{$\mathcal{G} = (\mathbf{V}, \mathbf{E}_d, \mathbf{E}_b)$: ADMG,\\
         $~~~~~~~~~~~~~$ $\mathbf{V}' \subseteq \mathbf{V}$: new Measure set ,
}
\KwOutput{Sub-ADMG where the observed nodes are changed to $\mathbf{V}'$ and new latent nodes are added to the current set of latent nodes.}

\vspace{0.2cm}
\hrule 
\vspace{0.2cm}

$\mathbf{U} = \mathbf{V} \backslash \mathbf{V}'$




            




\tcp{Apply simplification rules in \secref{sec:GraphNotCausalInf}}

$\mathcal{G'}$  = simplifiedNetwork($\mathcal{G}$, $\mathbf{U}$)
    
$\mathbf{return}$ $\mathcal{G'}$
        
\end{algorithm}

\clearpage

\onecolumn
\section{Including mediators into estimation of causal query reduce the asymptotic variance}
\medskip \textbf{Lemma 1} {\it Consider an ADMG A, where $T$ is the target of intervention and $Y$ is the effect. Including the mediators in any statistical estimator results in less variance than not including any of them.}

\begin{proof}
First, assume that there is only a single mediator $M_1$. The chain rule for mutual information implies that
\begin{eqnarray}
    I(T,M_1;Y) = I(T;Y) + I(M_1;Y|T)
\end{eqnarray}
The mutual information is non-negative. This implies that $I(T,M_1;Y) > I(T;Y)$, i.e., the amount of information obtained about $Y$ by observing both $T$ and $M_1$ is more than the amount of information obtained about $Y$ by observing only $T$. 
A higher mutual information value indicates a larger reduction of uncertainty whereas a lower value indicates a smaller reduction. Hence, measuring $M_1$ helps to reduce the uncertainty about $Y$, and it reduces its variation. 

Next, we extend this property to multiple mediators, $M_1, ..., M_{k}$. Expanding the chain rule for all the mediators, 
\begin{eqnarray}
    I(T, M_1, ..., M_{k}; Y) = I(T;Y) + I(M_1;Y | T)
                           + \sum_{i=2}^{k}I(M_i; Y | T, M_1, ..., M_{i-1}) \nonumber
\end{eqnarray}
Following the same logic as above, $I(T, M_1, ..., M_k; Y) > I(T,Y)$, i.e., the amount of information obtained about $Y$ by observing both $T$ and $M_1,..., M_{k}$, exceeds the amount of information obtained about $Y$ by observing only $T$.
\end{proof}

Lemma 1 implies that including all or a subset of the mediators improves the precision of the query estimation.

{\bf Causal query estimation details of the Case studies}

{\bf Case study 1}
For the causal generative approach, we 
modeled the exogenous variables with a Gaussian distribution.
The rest of the variables were modeled by representing the biomolecular reactions with a Hill function as described in Case study 2, Step 1. 
The non-informative $\mathcal{N}(0,10)$ priors were used for the parameters $\theta$ in the sigmoid function.
The linear regression in \figref{fig:IGFgenerateSetOfRankedSubADMGs} used the equations in \tableref{tables:LMFormulasCase1}.

{\bf Case study 2}
For the semi-and non-parametric approaches we directly used the \texttt{Ananke} library \cite{bhattacharya2020semiparametric}. For the linear regression approach we used the equations in \tableref{tables:LMFormulasCase2}.

\begin{table*}[t]
\centering
\begin{small}
\begin{tabular}{|c | c |} 
 \hline
\textbf{Figure}
& 
\textbf{Formula}
 \\[1mm]
 \hline
 \textbf{\figref{fig:IGFgenerateSetOfRankedSubADMGs} (a)} &

\begin{tabular}{c}
{} \\
   $Raf = \beta_0 + \beta_1 AKT + \beta_2 PI3K + \beta_3 Ras + \beta_4 SOS + \varepsilon_1$  \\
   {} \\
   $Mek = \beta_5 + \beta_6 Raf + \beta_7 AKT + \beta_8 PI3K + \beta_9 Ras + \beta_{10} SOS + \varepsilon_2$ \\
   {} \\
   $Erk =  \beta_{11} + \beta_{12} Mek + \beta_{13} Raf + \beta_{14} AKT + \beta_{15} Ras + \beta_{16} PI3K + \beta_{17} SOS + \varepsilon_3$\\
   {}
\end{tabular}
 \\

 \hline
 \textbf{\figref{fig:IGFgenerateSetOfRankedSubADMGs} (b)} &
 \begin{tabular}{c}
{} \\
$Erk = \beta_{18} + \beta_{19} AKT + \beta_{20} PI3K + \varepsilon_4$ \\
   {}
\end{tabular}
\\
\hline
 \textbf{\figref{fig:IGFgenerateSetOfRankedSubADMGs} (c)} &
 \begin{tabular}{c}
{} \\
$Mek = \beta_{21} + \beta_{22} AKT + \beta_{23} PI3K + \varepsilon_5$  \\
{} \\
$Erk = \beta_{24} + \beta_{25} Mek + \varepsilon_6$ \\
{}
\end{tabular}
\\
\hline
 \textbf{\figref{fig:IGFgenerateSetOfRankedSubADMGs} (d)} &
  \begin{tabular}{c}
{} \\
$Erk = \beta_{26} + \beta_{27} AKT + \beta_{28} SOS + \beta_{29} PI3K + \varepsilon_7$ \\
{}
\end{tabular}
\\
\hline
\end{tabular}
\end{small}
\caption{\small \textbf{Case study 1. Formulas used to estimate the query} $E[Erk | do(Akt = 80) ]$. The error terms ($\varepsilon_1, ..., \varepsilon_7$) are $i.i.d$ and follow $\mathcal{N}(0,\sigma^2_i)$ for $i \in \{1,...,7\}$ 
\label{tables:LMFormulasCase1}}
\end{table*}

{\bf Case study 3}
For the causal generative approach, we modeled the variables as in the causal generative approach in Case study 1.
For the linear regression approach, we used the equations in \tableref{tables:LMFormulasCase3}. 

\begin{table*}[b]
\centering
\begin{small}
\begin{tabular}{|c | c |} 
 \hline
\textbf{Figure}
& 
\textbf{Formula}
 \\[1mm]
\hline
 \textbf{\figref{fig:CovidCaseStudy} (a)} &
 \begin{tabular}{c}
 
 \begin{tabular}{c}
 \ {} \\
 $NFKB = \gamma_0 + \gamma_1 EGFR + \gamma_2 ADAM17 + \gamma_3 TNF + \gamma_4 SARSCOV2 $ \\
 $ + \gamma_5 ACE2 + \gamma_6 Ang + \gamma_7 AGTR1
+ \gamma_8 Gefi + \gamma_9 PRR + \epsilon_1$
 \end{tabular}
\\
{}\\
\begin{tabular}{c}
$IL6AMP = \gamma_{10} + \gamma_{11} NFKB + \gamma_{12} IL6STAT3 + \gamma_{13} ACE2
+ \gamma_{14} ADAM17$ \\
$+ \gamma_{15} TNF + \gamma_{16} EGFR + \gamma_{17} Toci
+ \gamma_{18} Gefi + \gamma_{19} PRR + \epsilon_2$ \\ 
\end{tabular}
\\
{} \\
\begin{tabular}{c}
$cytok = \gamma_{20} + \gamma_{21} IL6AMP + \gamma_{22} SARSCOV2 + \gamma_{23} ACE2 + \gamma_{24} Ang $ \\
$+ \gamma_{25} AGTR1 + \gamma_{26} Sil6r + \gamma_{28}TNF + \gamma_{29}NFKB + \gamma_{30}Gefi + \epsilon_3$  \\
\end{tabular}
\\
{}\\
\end{tabular}\\
\hline
 \textbf{\figref{fig:CovidCaseStudy} (b)} &
\begin{tabular}{c}
{} \\
$cytok = \gamma_{31} + \gamma_{32} EGFR  + \gamma_{33} ADAM17 + \gamma_{34} TNF$ \\
$+ \gamma_{35} Sil6r + \gamma_{36} IL6STAT3 + \epsilon_4$ \\
{} \\
\end{tabular}
\\
\hline
 \textbf{\figref{fig:CovidCaseStudy} (c)} &
\begin{tabular}{c}
{} \\
$cytok = \gamma_{37} + \gamma_{38} EGFR + \gamma_{39} ADAM17 + \gamma_{40} TNF + \gamma_{41} sIL6R\alpha$ \\
$+ \gamma_{42} IL6STAT3 + \gamma_{43} Gefi + \gamma_{44} EGF + \epsilon_5$ \\
{} \\
\end{tabular}
\\
\hline
 \textbf{\figref{fig:CovidCaseStudy} (d)} &
\begin{tabular}{c}

 \begin{tabular}{c}
 {}\\
 $NFKB = \gamma_{45} + \gamma_{46} EGFR + \gamma_{47} ADAM17 + \gamma_{47} TNF$ \\
 $+ \gamma_{48} ACE2  + \gamma_{49} Gefi + \epsilon_6$
 \end{tabular}
\\
{}\\
\begin{tabular}{c}
$IL6AMP = \gamma_{50} + \gamma_{51} NFKB + \gamma_{52} IL6STAT3 + \gamma_{53}ACE2 + \gamma_{54}ADAM17$ \\
$+ \gamma_{55}TNF + \gamma_{56} EGFR + \gamma_{57} Gefi + \gamma_{58}PRR + \epsilon_7$
\end{tabular}
\\
{} \\
\begin{tabular}{c}
$cytok = \gamma_{59} + \gamma_{60} IL6AMP  + \gamma_{61} ACE2 + \gamma_{62} Sil6r$ \\
$+ \gamma_{63} TNF + \gamma_{64} NFKB + \gamma_{65} Gefi + \epsilon_8$ \\
{} \\
\end{tabular}
\\
\end{tabular}  \\
\hline
\end{tabular}
\end{small}
\caption{\small \textbf{Case study 2. Formulas used to estimate the ATE}.
The error terms ($\epsilon_1, ..., \epsilon_8$) are $i.i.d$ and follow $\mathcal{N}(0,\sigma^2_i)$ for $i \in \{1,...,8\}$ 
\label{tables:LMFormulasCase2}}
\end{table*}

\begin{table*}[t]
\centering
\begin{small}
\begin{tabular}{|c | c |} 
 \hline
\textbf{Figure}
& 
\textbf{Formula}
 \\[1mm]
\hline
 \textbf{\figref{fig:experimentalCaseStudy} (a)} &
\begin{tabular}{c}
{}\\
\begin{tabular}{c}
$dpiA = \phi_0 + \phi_1 fnr + \phi_2 crp + \phi_3 rpoD + \phi_4 oxyR + \phi_5 dcuR + \phi_6 narL + \phi_7 fur$ \\
$+ \phi_8 aspC + \phi_9 dpiB + \phi_{10} hyaA + \phi_{11} appY + \phi_{12} hyaB + \phi_{13} hyaF + \varepsilon_1$ \\
{}\\
\end{tabular}
{} \\
$fnr = \phi_{14} + \phi_{15} fur + \phi_{16} rpoD + \phi_{17} crp + \varepsilon_2$ \\
{}\\
\end{tabular}  \\
\hline
 \textbf{\figref{fig:experimentalCaseStudy} (b)} &
\begin{tabular}{c}
{}\\
$dpiA = \phi_{18} + \phi_{19} fnr + \phi_{20} crp + \phi_{21} rpoD + \phi_{22} oxyR + \phi_{23} dcuR + \phi_{24} narL + \phi_{25} fur + \varepsilon_3$ \\
{} \\
$fnr =  \phi_{26} + \phi_{27} fur + \phi_{28} rpoD + \varepsilon_4$ \\
{}\\
\end{tabular}  \\
\hline
 \textbf{\figref{fig:experimentalCaseStudy} (c)} &
\begin{tabular}{c}
{} \\
$dpiA = \phi_{29} + \phi_{30} fur + \phi_{31} crp + \phi_{32} rpoD + \varepsilon_5$ \\
{} \\
\end{tabular}  \\
\hline
 \textbf{\figref{fig:experimentalCaseStudy} (d)} &
\begin{tabular}{c}
{} \\
$dpiA = \phi_{33} + \phi_{34} fnr + \phi_{35} fur + \phi_{36} rpoD + \phi_{37} dcuR + \phi_{38} crp + \varepsilon_6$ \\
{} \\
$fnr = \phi_{39} + \phi_{40} fur + \phi_{41} rpoD + \phi_{42} crp + \varepsilon_7$ \\
{} \\
\end{tabular}  \\
\hline
 \textbf{\figref{fig:experimentalCaseStudy} (e)} &
\begin{tabular}{c}
{}\\
$dpiA = \phi_{43} + \phi_{44} fnr + \phi_{45} fur + \phi_{46} rpoD + \phi_{47} dcuR + \phi_{48} narL + \phi_{49} crp + \varepsilon_8$\\
{}\\
$fnr = \phi_{50} + \phi_{51}fur + \phi_{52} rpoD + \varepsilon_9$\\
{}\\
\end{tabular} \\
\hline
\end{tabular}
\end{small}
\caption{\small \textbf{Case study 3. Formulas used to estimate the query $E[dpiA|do(fur=0)]$}.
The error terms ($\varepsilon_1,..., \varepsilon_9$) are $i.i.d$ and follow $\mathcal{N}(0,\sigma^2_i)$ for $i \in \{1,...,9\}$.
\label{tables:LMFormulasCase3}}
\end{table*}

\end{document}